\documentclass{aa}  

\usepackage[comma,authoryear]{natbib}
\usepackage{graphicx}
\usepackage[varg]{txfonts} 
\usepackage{lscape}

\usepackage{rotating}
\newcommand{\vsini}{$v\sin{i}$}
\newcommand{\eg}{{\it e.g.}}
\newcommand{\ie}{{\it i.e.}}
\newcommand{\cms}{cm\,s$^{\rm -1}$}
\newcommand{\ms}{m\,s$^{\rm -1}$}
\newcommand{\kms}{km\,s$^{\rm -1}$}
\newcommand{\kmts}{km$^{\rm 2}$\,s$^{\rm -1}$}

\newcommand{\ME}{M$_{\rm Earth}$}
\newcommand{\Msun}{M$_{\sun}$}

\newcommand{\harps}{H{\small ARPS}}
\newcommand{\espresso}{E{\small SPRESSO}}
\newcommand{\gclef}{G{\small -CLEF}}

\begin{document}

   \title{Using the Sun to estimate Earth-like planets detection capabilities.}

   \subtitle{V. Parameterizing the impact of solar activity components on radial velocities.}

   \author{S.Borgniet
          \and
           N.Meunier
          \and
           A.-M. Lagrange
          }

   \institute{
Univ. Grenoble Alpes, IPAG, F-38000 Grenoble, France\\ 
CNRS, IPAG, F-38000 Grenoble, France\\
\email{Simon.Borgniet@obs.ujf-grenoble.fr}
             }
   \date{Received ... ; accepted ...}

 
  \abstract
   {Stellar activity induced by active structures such as stellar spots and faculae is known to strongly impact the radial velocity (RV) time series. It is thereby a strong limitation to the detection of small planetary RV signals, such as that of an Earth-mass planet in the habitable zone of a solar-like star. In a serie of previous papers, we studied the detectability of such planets around the Sun observed as a star in an edge-on configuration. For that purpose, we computed the RV, photometric and astrometric variations induced by solar magnetic activity, using all active structures observed over one entire cycle.}
   {Our goal is to perform similar studies on stars with different physical and geometrical properties. As a first step, we focus on Sun-like stars seen with various inclinations, and on estimating detection capabilities with forthcoming instruments.}
   {To do so, we first parameterize the solar active structures with the most realistic pattern so as to obtain results consistent with the observed ones. We simulate the growth, evolution and decay of solar magnetic features (spots, faculae and network), using parameters and empiric laws derived from solar observations and literature. We generate the corresponding structures lists over a full solar cycle. We then build the resulting spectra and deduce the RV and photometric variations, first in the case of a ``Sun'' seen edge-on and then with various inclinations. The produced RV signal takes into account the photometric contribution of spots and faculae as well as the attenuation of the convective blueshift in faculae. We then use these patterns to study solar-like stars with various inclinations.}
   {The comparison between our simulated activity pattern and the observed one validates our model. We show that the inclination of the stellar rotation axis has a significant impact on the photometric and RV time series. RV long-term amplitudes as well as short-term jitters are significantly reduced when going from edge-on to pole-on configurations. Assuming spin-orbit alignment, the optimal configuration for planet detection is an inclined star ($i \simeq 45\degr$).}
   {}

   \keywords{Stars : activity -- Stars : starspots -- Stars : solar-type -- Techniques : radial velocities -- Techniques : photometric}

   \maketitle
%

\section{Introduction}
Thanks to the radial velocity (RV) technique, more than 500 exoplanets have been discovered in the two last decades (http://exoplanets.eu). First limited to Jupiter-like planets on short period orbits, the RV technique allows today to detect Neptune-like planets (10-40 \ME) and Super Earths (1.2-10 \ME) on longer period orbits. This progress was made possible thanks to the important improvements on the sensitivity and stability of instruments, as well as observational strategies to average out known sources of stellar signals \citep{dumusque11b}. For example, the High-Accuracy Radial velocity Planet Searcher (\harps) spectrograph located on the ESO 3.6 m telescope in La Silla \citep{pepe02} gives a precision of 1 \ms~for a Solar-type star in average conditions, compared to 5-10 \ms~with previous instruments. Furthermore, future instruments, such as \espresso~on the Very Large Telescope (VLT) or \gclef~on the Giant Magellan Telescope (GMT) are expected to reach a precision down to 0.1 \ms~\citep{megevand10,frez14}. Such a level of precision will theorically give access to lower planetary masses far from their host stars such as Earth-like planets in the habitable zone (hereafter HZ). 

 However, low-amplitude RV planet signals (such as RV signatures of Earth-mass planets in the HZ) are much more sensitive to stellar perturbations than giant planet RV signatures. In the case of a low-mass planet RV signature with an amplitude in the 0.1-1 \ms~range, the stellar noise or ``jitter'' is high enough to either mask or mimic the planet-induced RV variations, even in the case of a chromospherically nearly inactive star \citep{isaacson10,lagrange10,meunier10a}.

 The so-called stellar jitter mainly comes from three different sources: stellar oscillations or pulsations, granulation, and stellar magnetic activity.

 Stellar pulsations or oscillations dominate the RV jitter on the shorter timescales. For FGK dwarfs, they are mainly driven by accoustic or pressure waves ({\it p}-modes). Pressure waves are commonly believed to originate from turbulent convective motions occuring in the stellar outer layers and propagate through the star. They induce RV shifts with an amplitude of a few \cms~to one \ms. The oscillation periods range from a few minutes (\eg, five minutes in the case of the Sun) to a few tens of minutes. The amplitude and the period of the oscillations increase with stellar mass. Long exposure times and observational strategies have shown to be efficient in averaging the RV noise induced by pulsations in the case of late-type stars \citep{santos04,dumusque11a}.

 The photospheric granulation phenomenon accounts for the convective plasma motions occuring in the outer envelop of solar-mass stars. The quick rise and fall of the plasma results in a pattern of bright granules and darker lanes at the stellar surface, and in rapidly evolving RV shifts, with amplitudes up to a few \kms~locally. When integrated over the entire stellar disc, upflows and downflows average out, leaving a residual RV jitter at the level of the \ms. Up to now, there have been few tentatives to estimate the impact of granulation on the RV. \cite{dumusque11a} made a first estimation for different spectral type stars based on the study of asteroseismology measurements. \cite{cegla13} developped a four-component model of granulation, building absorption line profiles from three-dimensional magnetohydrodynamic solar simulations. This study was continued by \cite{cegla14}, who derived the corresponding RV time series as well as other time series of observables such as the bisector amplitude. They concluded first that granulation has a strong impact on RV at the \cms~to \ms~levels and would be a potentially significant limitation to low-mass planets detection. Second, they found correlations between the RV time series and other observables (\eg~bisector curvature or bisector inverse slope) that already allow to partially correct the granulation signal and reduce the corresponding jitter by one third. Meunier et al ({\it submitted}) simulated a collection of both granules and supergranules on the solar visible hemisphere over a whole solar cycle and derived the induced RV and photometric time series. These authors concluded that the granulation noise is hard to average out even over an entire night and will therefore have a significant impact on detection limits.

 From a few days to several years timescales, the stellar jitter is dominated by the so-called stellar magnetic activity. Active regions such as dark spots an bright faculae (colder and hotter than the quiet photosphere, respectively) grow and evolve on the stellar surface, inducing changes in the stellar irradiance and RV variations (as the flux loss or excess in the active regions distorts the CCF of the stellar spectra). The stellar rotation makes one see these active regions moving across the stellar disc, thus inducing an apparent Doppler shift. The first studies of the impact of starspots on RV and line bisectors were made by \cite{saar97} and \cite{hatzes02}. \cite{saar97} built a semi-empirical law that directly bound the spot-induced RV semi-amplitude, the spot filling factor and the stellar rotational velocity for cool F-G stars. \cite{saar03,saar09} made similar studies for bright faculae. The first spot model was developed by \cite{desort07}. The authors computed synthetic stellar spectra and applied a black-body law to take into account the contribution of a colder spot. They quantified accurately the RV amplitude, RV bisector and photometric variations induced by a starspot for a range of different stellar and spot properties.

Such simple spot models have proved to be useful and efficient to simulate the RV signature of one or a few active regions during a timescale of the order of the stellar rotational period. They are well adapted to stars that host a single main spot or a few ones, such as active young dwarf stars \citep[\eg~\object{HD\,189733}, see][and ref. therein]{dumusque14b}. They are however not sufficient to reproduce the total activity-induced RV signal for Sun-like stars, due to the presence of multiple evolving active regions on the stellar disc, and due to the strong contribution of faculae because of slow rotation \citep{meunier10a,dumusque14a}. A complete model of the stellar activity pattern for longer timescales (of the order of the activity cycles) is needed. For most stars though, we have no detailed information on the activity properties as active structures are not directly observable. Doppler and Zeeman-Doppler imaging technics allow to recover the largest active structures (or active structure clusters) only for fast-rotating stars (\ie, young, active ones) but not for old, slow-rotating solar-type stars \citep{rice02}. The structure properties, as well as the convection ones, are mostly unknown and we rely only on indirect and global estimators such as the Calcium (Ca) index \citep[see \eg~][]{baliunas83,noyes84,hall04}.

The Sun is an exceptional star, for which we have a lot more of information. Indeed, both solar dark spots and bright features are well observed and the various solar activity properties are thus well described. The Sun, if seen as a moderately active star, therefore represents an ideal prototype to study the impact of activity on low-mass planets detectability. In a series of papers, we modelled with a great accuracy the solar activity pattern using detailed observations (\ie, spots catalogs and magnetograms) extending over the full solar Cycle 23. We then rebuilt the induced photometric variations and the corresponding spectra, and deduced from the latter the RV time series. We first considered only cold spots in \cite{lagrange10} (hereafter Paper I), studying their impact on the detectability of an Earth-mass planet in the HZ of a Solar-like star. In \cite{meunier10a} (hereafter Paper II), we extended this study by considering in addition the contribution of bright faculae and, for the RV, the attenuation of the convective blueshift in magnetically active regions. Combining the effect of the three activity components, we found out that the attenuation of convective blueshift in faculae (hereafter the convective component) dominated the induced RV variations (this happens as the Sun is a slow rotator: for larger \vsini, the photospheric effect of spots becomes dominant). We showed that unless with correction tools (still to be identified and tested) it would be impossible to detect an Earth-like planet in the HZ. Using the same simulation as in Paper II, \cite{lagrange11} (hereafter Paper III) estimated the astrometric effect of stellar magnetic activity. Finally, in \cite{meunier13} (Paper IV), we used the strong correlation between the convective component and the Ca index (as both are directly related to the chromospheric plage or photospheric facula filling factor) as a tool to correct RV time series from the convection-induced jitter. We showed that an Earth-like planet in the HZ would become detectable given that: {\it i}) there is an excellent signal-to-noise ratio (hereafter S/N) on the Ca index data; and {\it ii}) the temporal coverage of the RV observations over one activity cycle is very dense (typically one night out of four during the cycle).

\vspace{5mm}

The objective of the present paper is to extend the previous study to a Sun-like star seen under any inclination. To do so, we fully parameterize the activity pattern of a Sun-like star. A few activity parametrizations have already been developed. \cite{barnes10} undertook the first tentative to parameterize dark spot distributions at low and high activity levels in the case of M-type stars. A similar simple model was used by \cite{jeffers14}, this time for young and active G and K dwarfs. A parametrization of an activity pattern was made by \cite{dumusque11b}, who simulated dark spot distributions over the stellar surface for different levels of activity of a solar-like star. The authors derived detection limits for different activity levels and observational strategies. This study was however limited to dark spots and did not include the flux effect of bright features (faculae and network), nor the inhibition of the convective blueshift in active structures. 

Using the Sun as a template presents several important advantages: {\it i}) we have much more information on solar activity than for any other star; {\it ii}) the validity of such a model can be easily asserted by a comparison with the previous papers; and {\it iii}) such an activity model can be adapted to other spectral types and other activity levels than the Sun, if the active region configuration is supposed similar. In addition, it allows studying the impact of stellar inclination on the results, contrary to the solar observations for which only one hemisphere is observed at any given time.

 In Sect.~\ref{simus}, we build the spot and facula distributions on the Sun surface over a full solar cycle. We describe in details the laws and parameters we used to make these simulations. To test our approach, we then compare the obtained structure distributions with the observed solar distributions used in paper II. In Sect.~\ref{results}, we compute the induced RV variations, taking into account both the photometric contribution of spots and faculae and the attenuation of the convective blueshift in active structures. Then we compare our results with those previously obtained using observed solar magnetic features. In Sect.~\ref{inclin}, we study the influence of the stellar inclination on the resulting RV and photometry. The astrometric time series will be treated in a separate paper. We also compute the corresponding detection limits. We finally discuss our results and the use of our model for other stars in Sect.~\ref{conclu}. 


\section{Building the activity pattern} \label{simus}
 
\subsection{Approach}
We consider the solar activity pattern during Cycle 23. We use the Sun as a template and as a mean of comparison with the observed activity pattern used in Papers I and II. We parameterize an extended range of activity levels and timescales. We start from the general distribution of the activity level over the cycle, then we parametrize the spatial and temporal distributions of active structures and their dynamics, and we finally modelize the individual behaviour of the structures, including the dark spots, the bright faculae around these spots and the network smaller features. The two outputs of our simulations are the lists of the dark spots and of the bright features, respectively. Each file gives, for each time step of the simulation, the structure sizes (in millionth of hemisphere, hereafter $\mu$Hem), latitudes and longitudes. For a given set of parameters, these files will then represent the inputs of our simulation tool to produce the spectra and the corresponding RV at each time step, as it was done in Papers I and II with observed solar structure lists.

\subsection{Building the spot and facula catalogs: input parameters}\label{listparam}
All input parameters are summarized in Table.~\ref{parameters}. We detail them hereafter.
\begin{enumerate}
\item {\it Global activity level.} The sunspot-induced solar activity follows a cycle of $12 \pm 1$ years. To mimic as accurately as possible the global activity level and its distribution over such a cycle, we used the relative sunspot number, or Wolf number (noted $R$), as a proxy. For this study, we based our simulations on solar Cycle 23. This allowed us both to build a realistic activity pattern and to compare our results with those previously obtained in Paper II. The daily Wolf number $R$ for Cycle 23 was recovered at the Solar Influences Analysis Data Center\footnotemark~(SIDC) and is displayed in Fig.~\ref{wolfnbr} (black dots, upper panel). 
\footnotetext{available at : http://sidc.oma.be/ .}
We first smoothed the daily $R$ data to obtain a long-term reference $R_{\rm smth}$ (Fig.~\ref{wolfnbr}, solid line, two top panels) for the cycle shape. 

   \begin{figure}[ht!]
   \centering
   \includegraphics[width=1\hsize]{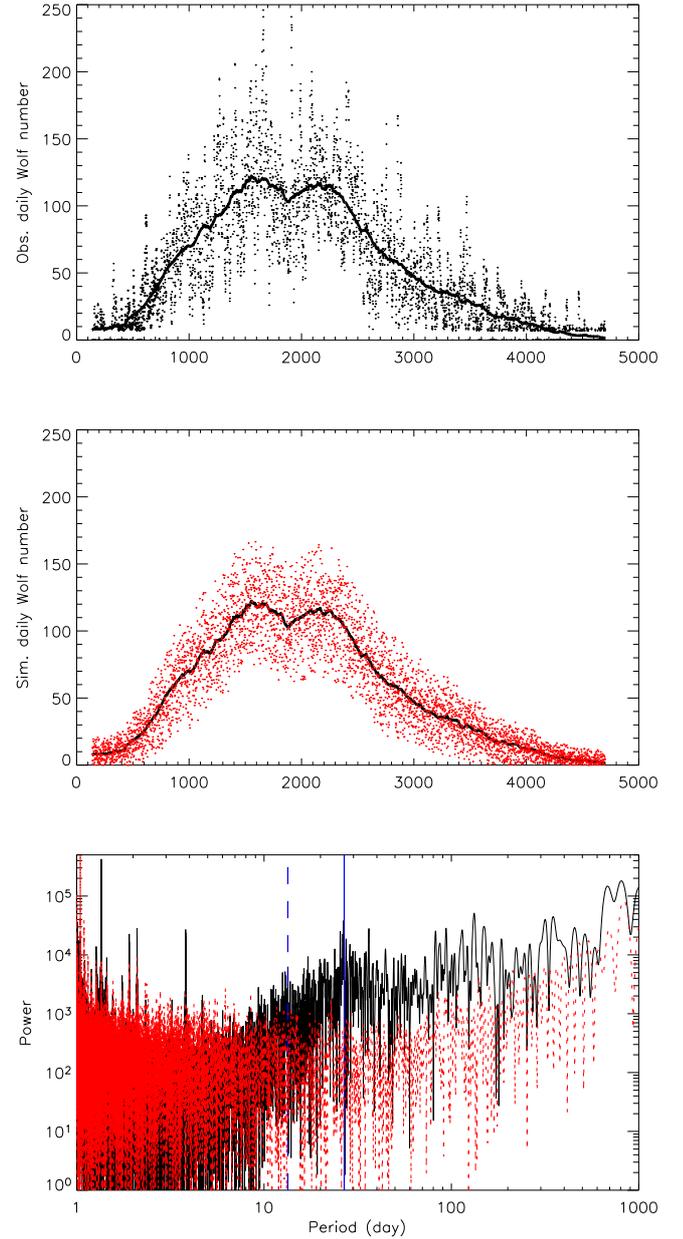}
   \caption{{\it Top}: daily ({\it dots}) and smoothed ({\it bold line}) Wolf number over solar cycle 23. {\it Middle}: same smoothed Wolf number ({\it black line}), and when randomly adding a second order polynomial dispersion ({\it red dots}). {\it Bottom}: Power spectra of the observed daily ({\it black}) and simulated ({\it red}) Wolf numbers (respectively $R$ and $R^{'}$). The solar rotation period ({\it blue solid line}) and its first harmonic ({\it blue dashed line}) are also displayed for comparison.}
   \label{wolfnbr}
   \end{figure} 

The observed daily dispersion of the Wolf number comes from two contributions: the first comes from the rotational modulation and will naturally be present in the model (hence the smoothing); the second is due to the fact that appearing spots do not follow a smooth curve and present some dispersion originating in the dynamo process. To take this contribution into account, we calculated a second order polynomial dispersion $d$ to the smoothed curve $R_{\rm smth}$, expressed as:
\begin{equation}
d = d_{0} + d_{1}R_{\rm smth} + d_{2}R_{\rm smth}^{2}
\end{equation}
where $d_{0}$, $d_{1}$, $d_{2}$ are derived empirically from observations. We then randomly added this dispersion $d$ to the smoothed data $R_{\rm smth}$ to obtain a daily input Wolf number $R^{'}$ (Fig.~\ref{wolfnbr}, red dots, middle panel). The Wolf number ($R$, $R^{'}$) is a combination of the number of individual spots and of the number of spot groups. As the input of our simulation is the total number of spots $N$ (which accounts both for the individual ones and for those in groups), we have derived a calibration to deduce this total number of spots from an activity level expressed in Wolf number units:
\begin{equation}\label{wolfeq}
 N = n_{0} +n_{1}R^{'}
\end{equation}
where $n_{0}$, $n_{1}$ are derived empirically from observations. Using Eq.~\ref{wolfeq}, we obtained the theoretical total number of sunspots expected on the solar surface at each time step. The conversion $n_{0,1}$ and dispersion $d_{0,1,2}$ coefficients used in our simulations are given in Table.~\ref{parameters}. By subtracting the number of sunspots already existing at this step, one can finally determine the number of new sunspots to be generated at this step in the simulation.\\

We finally display in Fig.~\ref{wolfnbr} (bottom panel) the power spectra of the observed $R$ and simulated input $R^{'}$ Wolf number for comparison purpose. The periodograms are quite different, especially at the rotational period and its first harmonic. This is expected as we do not introduce the rotational modulation (which is naturally present in the observed Wolf number) in our input Wolf number $R^{'}$ but rather at the next step of the simulations (see above). Concerning the high-amplitude (and short-period) variations of the observed Wolf number in the high activity solar period, we consider that these variations are not due to a lack of spots at some moments in our model compared to the observations, but rather to the presence at these times of a group of large spots or an active cluster in the solar observations (see below for the spot size and spot filling factor distributions). Our model would require a more complex spot distribution function to reproduce such time-to-time appearances of large spot groups and we decided not to complicate it at this stage. \\

We assumed an activity cycle length of 12.5 years (length of solar Cycle 23) and a time step of one day, leading to a total number of 4566 iterations or days. Furthermore, we added a random scattering of the time steps, with an amplitude of 4 hours around the regular daily period, to mimic real observations.\\

\item {\it Spatial and temporal distributions of the structures}. 
At the beginning of the activity cycle, sunspots appear at medium latitudes, with some dispersion. As the activity level increases, the mean appearance latitude of sunspot slowly decreases in absolute value. This migration of active regions towards the equator continues during the whole cycle leading to a low mean appearance latitude at the end of the cycle. In our simulations, the mean latitudes of sunspot appearance at the beginning and at the end of the cycle were fixed at $\pm $22\degr~and $\pm$ 9\degr, respectively. These values are derived from the spot catalogs we used in Papers I and II. A linear law was then used to fit the decrease of the sunspot appearance latitude during the cycle, and we added a latitude scattering to reproduce the butterfly diagram.\\

We also had the possibility to create an asymmetry between the activity patterns in the northern and southern hemispheres, \ie~by adding more spots in one of them. As such a significant asymmetry has not been detected on the Sun during past solar cycles, we decided to put the same proportion of spots on each hemisphere.

Spot groups are finally well-known not to appear at random longitudes, but around the so-called active longitudes. Such preferred longitudes have been detected on the Sun as well as on several other stars \citep{berdyugina03,ivanov07,lanza09,lanza10}. Their origin is still partly unexplained and is probably due to the dynamo process. Some trends seem to emerge from observations : there are generally two (seldom three) persistent active longitudes per hemisphere, shifted by $180\degr$. They form a rigid structure, although they cannot be fixed in a reference frame because of the differential rotation \citep{berdyugina03}. 

On the Sun, magnetic field and new activity seem to emerge at locations where activity is already present \citep[see \eg][]{harvey93}. Thus, instead of putting active longitudes with fixed longitude values in our model, we decided to build active longitudes from the position of already present sunspots. A fraction of new sunspots then appears in a restricted area in longitude around already present sunspots, while the remnant fraction is uniformly distributed in longitude. The fraction of new sunspots appearing in theses active areas and the longitude extension of the latter are reported in Table.~\ref{parameters}.\\

\item {\it Large scale dynamics}. We describe here the global motion of magnetic features at the solar surface during their lifetime. This motion can be decomposed in two components, namely a longitudinal and a latitudinal motion.\\
 The Sun is well known to exhibit a differential rotation in latitude: active structures rotate faster at the equator than at the poles. The longitudinal motion or rotation rate of the active structures $\omega$ (in degree per day) is therefore a function of the latitude $\theta$, according to the following equation \citep[\eg ][]{ward66,meunier05c}: 
\begin{equation}
\omega = \omega_{0} + \omega_{1}.sin^{2}(\theta) + \omega_{2}.sin^{4}(\theta)
\end{equation}
The latitudinal motion or meridional flow $M$ of magnetic features is derived from \cite{komm93}. It is a poleward motion in each hemisphere, which is also a function of the latitude :
\begin{equation}
M(\theta) = \alpha.sin(2\theta) + \beta.sin(4\theta)
\end{equation}

  \begin{table*}[t!] 
    \caption{Input parameters. Free parameters are displayed in italic.}
    \label{parameters}
    \begin{center}
      \begin{tabular}{l l l c c}
        \hline
	\hline
        &   Parameter  &  Value & Unit  & Reference \\
	\hline
        \hline
Solar activity     & Reference cycle  & Solar Cycle 23  &    & \\
        &   Cycle length            &  12.5    & [year]      & \\ 
        &   Time step               &  1       & [day]       & \\  
        &   Time step random dispersion & 4    & [hour]      & \\  
        &   Wolf number normalization   & $n_{0}$ = 1.32    &             &   \\   
        &                               & $n_{1}$ = 0.148   &             &   \\   
        &   Wolf number random dispersion & $d_{0}$ = 6.922   &             &  \\  
        &                                 & $d_{1}$ = 0.7594  &             &  \\ 
        &                                 & $d_{2}$ = -0.00348 &            &  \\  
        \hline
        \hline
Spatio-temporal     &  Mean start latitude     & $\pm 22$ &[$\degr$] & \\ 
 distribution       &  Mean end latitude       & $\pm 9$  &[$\degr$] & \\
                    & Standard lat. dispersion & 6 &[$\degr$] &        \\ 
                    & Max. lat. dispersion     & 20 & [$\degr$] &     \\  
                    & North/South asymmetry    & 0.5 &    -      &    \\  
                    & Active longitude spot fraction & 0.4 & -   &    \\  
                    & Active longitude extension area & $\pm 20$ &[$\degr$] &  \\  
        \hline
        \hline
Large scale dynamics & Spot differential rotation &  $\omega_{0} = 14.523$   &   [$\degr$/day]               &\cite{ward66}  \\ 
                     &                                  & $\omega_{1} = -2.688$     &  [$\degr$/day]                  & This paper.\\
                     &                                  &  $\omega_{2} = 0$       &   [$\degr$/day]                 & This paper.  \\
                     & Facula and network               & $ \omega_{\mathrm{b0}} = 14.562$ & [$\degr$/day]                      & \cite{meunier05c} \\
                     & differential rotation            & $ \omega_{\mathrm{b1}} = -2.04$& [$\degr$/day]                        & This paper. \\
                     &                                  &  $\omega_{\mathrm{b2}} = -1.49$& [$\degr$/day]                        & This paper. \\
                     & Meridional flow, all structures     & $\alpha = 12.9$ &    [\ms]          &\cite{komm93} \\  
                     &                                     &$ \beta = 1.4$   &    [\ms]          &  This paper.   \\  
                     & Stellar radius                      &$ 1 R_{\mathrm{\sun}} = 696400$ & [km]      &          \\   

        \hline
        \hline
Spots properties    &                 &     &  & \\ 
                    &  Isolated spots &     &  & \\ 
        \hline      
                    & Total fraction  & 0.4  &- &\cite{martinez93}    \\   
                    & Mean initial size &  46.51\tablefootmark{a} & [$\mu$Hem]  &\cite{baumann05}\tablefootmark{b} \\       
                    & Standard size deviation & 2.14  & [$\mu$Hem]  & This paper. \\
                    & Max. size         & 1500 & [$\mu$Hem]         &  Papers I and II         \\  
                    & Mean decay        & -18.9 &  [$\mu$Hem/day] &\cite{martinez93}\tablefootmark{c}\\ 
                    & Median decay      & -14.8 &  [$\mu$Hem/day] & This paper. \\ 
        \hline
                    &  Complex spot groups & & & \\
        \hline  
                    & Total fraction  & 0.6  &- &\cite{martinez93}     \\      
                    & Mean initial size &  90.24\tablefootmark{a}  & [$\mu$Hem]  &\cite{baumann05}\tablefootmark{d}\\   
                    & Standard size deviation & 2.49  & [$\mu$Hem]  & This paper.   \\
                    & Max. size         & 5000 & [$\mu$Hem]       & Papers I and II           \\  
                    & Mean decay        & -41.3 &  [$\mu$Hem/day] &\cite{martinez93}\tablefootmark{e}  \\
                    & Median decay      & -30.9 &  [$\mu$Hem/day] & This paper.  \\ 
        \hline
                    & Both spot types     & & &    \\ 
        \hline
                    & Min. decay value  & -3    &  [$\mu$Hem/day] &             \\   
                    & Max. decay value  & -200    &  [$\mu$Hem/day] &            \\ 
                    & Min. spot size    & 10   & [$\mu$Hem]        & Papers I and II \\     
        \hline
        \hline
Faculae properties  & {\it q (facula-to-spot ratio) } &  &               & \\ 
                    & {\it Mean log(q) }  & 0.8 &- &               \\    
                    & {\it Standard deviation (log(q)) } & 0.4 &- &     \\   
                    & {\it Min.- Max. log(q)     }        & 0.1 -- 5 &- & \\  
                    & Mean decay                    & -27      & [$\mu$Hem/day] & \\ 
                    & Median decay                  & -20      & [$\mu$Hem/day] &  \\
                    & Min. facula size              & 3        & [$\mu$Hem]      & Papers I and II \\ 
        \hline
        \hline            
Network properties  &  Diffusion coefficient        & 300      & [\kmts]  &\cite{schrijver01}\\ 
                    &  {\it Remainder fraction for decay} & 0.975    & [ /day]                &  \\  
                    & Min. size                     & 3        & [$\mu$Hem]           & Papers I and II \\ 
                    & {\it Facula fraction recovered in network} & 0.8 & -               &               \\      
        \hline
        \hline
      \end{tabular}
 \tablefoot{
\tablefoottext{a}{ For these two parameters, the value taken from \cite{baumann05} was multiplied by a factor 1.54 to better fit the spot catalogs used in Papers I and II.}
\tablefoottext{b}{see the Table 1. of the paper, total area of single spots, snapshot method.}
\tablefoottext{c}{Table 3., standard cycle.}
\tablefoottext{d}{Table 1., total area, snapshot method.}
\tablefoottext{e}{Table 1., standard cycle.}
}
    \end{center}
  \end{table*}

\item {\it Spot properties}.  Spots appear preferentially with a given size, increase in area rapidly before slowly decreasing. The spot growing phase is very quick \citep[in average 10 to 11 times faster than the decay phase, see][]{howard92a}. It is then often shorter than our one-day timestep for the small and mid-sized sunspots and of a few days for the largest ones. As we focus here on long-period planets, and as the active structure growing phase is smaller or of the order of our simulation time step, we decided not to include it in our model for the moment and assumed that spots appear with their maximal size and then decrease during all their lifetime. We will however include a description of the growing phase in future works to better describe the variations at small timescales. Thus the spot distribution can be described with only the initial size distribution and a decay law. According to the La Laguna classification \citep[see][and ref. therein]{martinez93}, one can distinguish between isolated spots (La Laguna type 3) and spots belonging to complex groups (La Laguna type 2). In compliance with this classification, we built two distinct spot distributions. Both obey to similar evolution laws, but with different parameters. The initial spot size distribution can be fitted with a log-normal law, according to \cite{baumann05}. The two input parameters are therefore the mean spot size and the standard spot size deviation. We took the parameters used in our simulations from the snapshot model developed in \cite{baumann05}. The spot decay law has been parametrized by \cite{martinez93} as a log-normal law, with two input parameters~: the mean and median decay values. We also put an upper size limit for the two types of spots, and a lower threshold of 10 $\mu$Hem for all spots, chosen to be in good agreement with the observed spots distribution used in Paper II.\\
    
\item {\it Facula properties}. On the Sun, spots and faculae form active regions, spots being surrounded by large faculae. The ratio between the surface covered by faculae and spots (hereafter the size ratio) has been studied \citep{chapman97,chapman01,chapman11}. The authors found an average size ratio ranging between about 13 and 45 over solar cycles 22 and 23, depending on the method used to estimate the facular contrast. However, this corresponds to an average ratio measured for structures that can be at any state of evolution and any size. As for our simulation, we need instead an initial size ratio. Indeed, each time we add a spot, we add a facula at the same place. The facula size distribution then depends only on the initial size ratio. As there is little in the literature about this quantity, it is one of our few free parameters. We assume a log-normal law to describe the initial facula-to-spot size ratio. Facula decay is then described by two processes:
\begin{itemize}
\item First the ``classical'' decay is described as for dark spots, with a log-normal law. This decay corresponds to a certain surface being lost by the facula at each time step of the simulation.
\item Second, a given proportion of this facula lost surface is converted into network (the rest being completely lost, for example by flux cancellation or submergence): facula fragments break away from active regions and diffuse over the surface.
\end{itemize}
As the facula distribution is based on the spot distribution, the facula growing phase is not reproduced for the moment in our model. The facula growth and decay are longer than the spot ones in average, due to their longer lifetimes \citep{howard92a}, but the facula growth is still much shorter than its decay \citep[see \eg][]{howard91}, of a few days then. We decided not to reproduce it to remain coherent with the spot distribution and as we focus mainly on timescales between a fraction of the rotational period and a complete activity cycle, and on long-period planets.\\
 

\item {\it Network properties}. We chose to make the network originates in the decay of faculae (see point 5). Network behavior is then managed by its diffusion coefficient and by its decay rate. The network diffusion coefficient has been studied in many papers which provide a great range of values \citep[][and ref.therein]{cadavid99}.

   \begin{figure}[ht!]
   \centering
   \includegraphics[width=1\hsize]{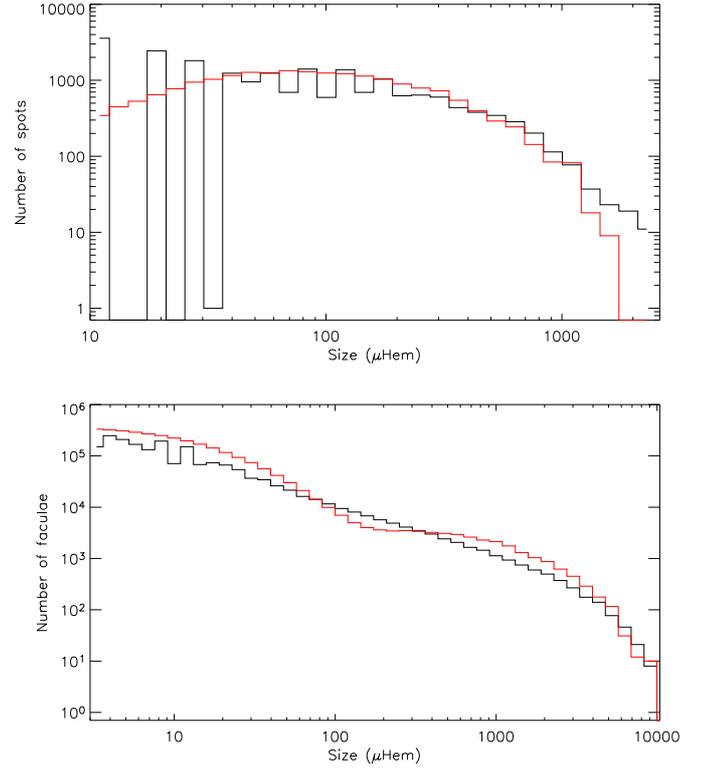}
   \caption{{\it Upper panel}: observed (black line) and simulated (red line) spot sizes. {\it Lower panel}: same for bright feature sizes.}
   \label{histo_size}
   \end{figure} 

As these authors pointed out, there is a large gap between the diffusion coefficient values found by modeling (which are usually higher than 500 \kmts) and the values found with magnetograms analysis (which span from a few tens of \kmts~to less than 300~\kmts). They also raised out the possibility for the diffusion coefficient to be time-dependent.\\

We finally decided to keep the value of 300 \kmts~used by \cite{schrijver01} as it stands well between the two extreme groups of values. We assume the diffusion process to be isotropic \citep{cadavid99}. As for the decay rate, we simply described it by fixing the network remainder fraction at each time step (it is one of the model free parameters). We finally put a minimal size of 3 $\mu$Hem for all bright structures (faculae or network), again in compliance with Paper II.

\end{enumerate}

\subsection{Comparison with the observed activity pattern}\label{comps}

Here we compare the structures generated with our parameterized model to the observed structures used in Paper II. To do this, we use the spot and bright feature lists over Cycle 23 that we retrieved from sunspot catalogs and MDI/SOHO magnetograms, respectively (Paper II). These data sets have some gaps, leading to a coverage of 3586 days, for a total duration of 4171 days, whereas our simulations have a length of 12.5 years with a daily sampling, \ie~4566 days (length of Cycle 23). We therefore applied to our outputs the same calendar as in Paper II so as to make a relevant comparison. The full time series will be studied in Sect.~\ref{inclin}.

\subsubsection{Size distributions}\label{sizedist}

   \begin{figure*}[ht!]
   \centering
   \includegraphics[width=0.9\hsize,height=0.55\hsize]{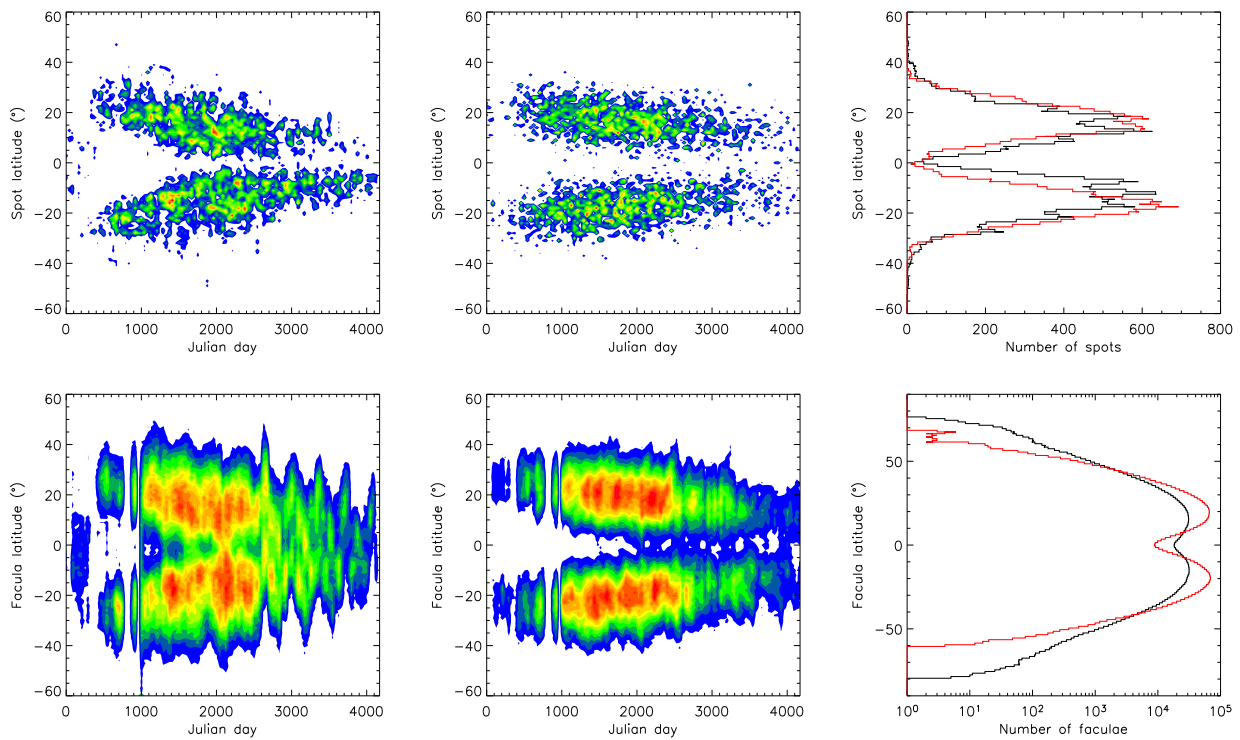}
  \caption{\footnotesize{{\it Upper panel}: observed ({\it left}) and simulated ({\it middle}) spot latitude distribution over one Solar cycle. {\it Right}: Histogram of the observed (black) and simulated (red) spot latitude distributions. {\it Lower panel}: same for all bright features (faculae and network). Note that in the case of bright features, the histogram has its {\it x}-axis in log-scale.}} 
   \label{histo_lat}
   \end{figure*}

The size distributions (in $\mu$Hem) of our simulated structures are displayed in Fig.~\ref{histo_size} and compared with the observed ones from Paper II. We want to highlight the fact that very few of the input parameters detailed in Table~\ref{parameters} are free parameters. For most of them, we retrieved the input laws and parameters from the literature so as to build a fully parameterized model of solar-like activity.\\

 For the dark spot properties, we have no free parameters (\ie, they all originate from the literature). Remarkably, we find a good agreement between the spot size distribution simulated this way and the observed one. The lack of a few large spots (about 40 in all) for the simulated distribution is explained by the fact that the log-normal law usually used in the litterature slightly underestimates the number of very large spots \citep[see also Figs. 2 and 3 in][]{baumann05}. The short-term differences reported above in the Wolf number distributions are then likely to originate in the time-to-time appearance of these very large spots (and not to a global lack of spots in our model, as the simulated distribution fits well the observed one). As for smaller spots, the observed distribution is discretized, which partly explained the gap (the remaining difference is again due to the fact that the observed distribution slightly differs from the log-normal law). This small discrepancy in the distribution for small spots will not influence significantly the RV signal as the RV effect of small spots is negligible compared to the larger ones.\\

For the bright features (\ie, faculae and network, hereafter), there are three free parameters (the facula-to-spot initial size ratio distribution, the facula fraction recovered in network and the network decay rate). These parameters are poorly documented, so we decided to adjust them to better fit the observed distribution. Overall, the simulated size distribution is in good agreement with the observed one (Fig.~\ref{histo_size}). However, there are some remaining differences which are inherent to the model parameters. The size distribution for all observed bright features does not follow a log-normal law. For the simulated bright features, we remind that we chose to make the facula initial size distribution directly dependent on the spot distribution, and the network originate in the decrease of the faculae. This explains thus the ``flattened-S'' shape of the simulated distribution, which corresponds to the addition of the facula and network respective distributions. Reproducing even more accurately the shape of the observed distribution would require to make our model much  more complex, for example by making the bright feature decay size-dependent or cycle-dependent. Given our goals, such a level of complexity is not necessary. As larger faculae have the largest influence on the RV, we decided to fit the distribution in priority for the highest sizes (\ie, for sizes greater than 5000 $\mu$Hem) by adjusting our free parameters. There is then still a discrepancy between the observed and simulated distributions for the smallest sizes, but we consider that its effect on our observables will be mostly negligible at our level of precision.

\subsubsection{Latitude distributions}

Another way to compare our simulated pattern to the observed one is to study the active structure latitude distributions. The latter are displayed in Fig.~\ref{histo_lat}. By doing so we will be able to validate the large scale behavior of our model. The latitude distributions of observed and simulated structures are indeed in good agreement. The kind of oscillations that we can distinguish in the observed bright feature latitude distribution at high latitudes and with a 1-year period comes from a ``seasonal'' effect. 

  \begin{figure*}[ht!]
  \centering
\includegraphics[width=0.9\hsize,height=0.52\hsize]{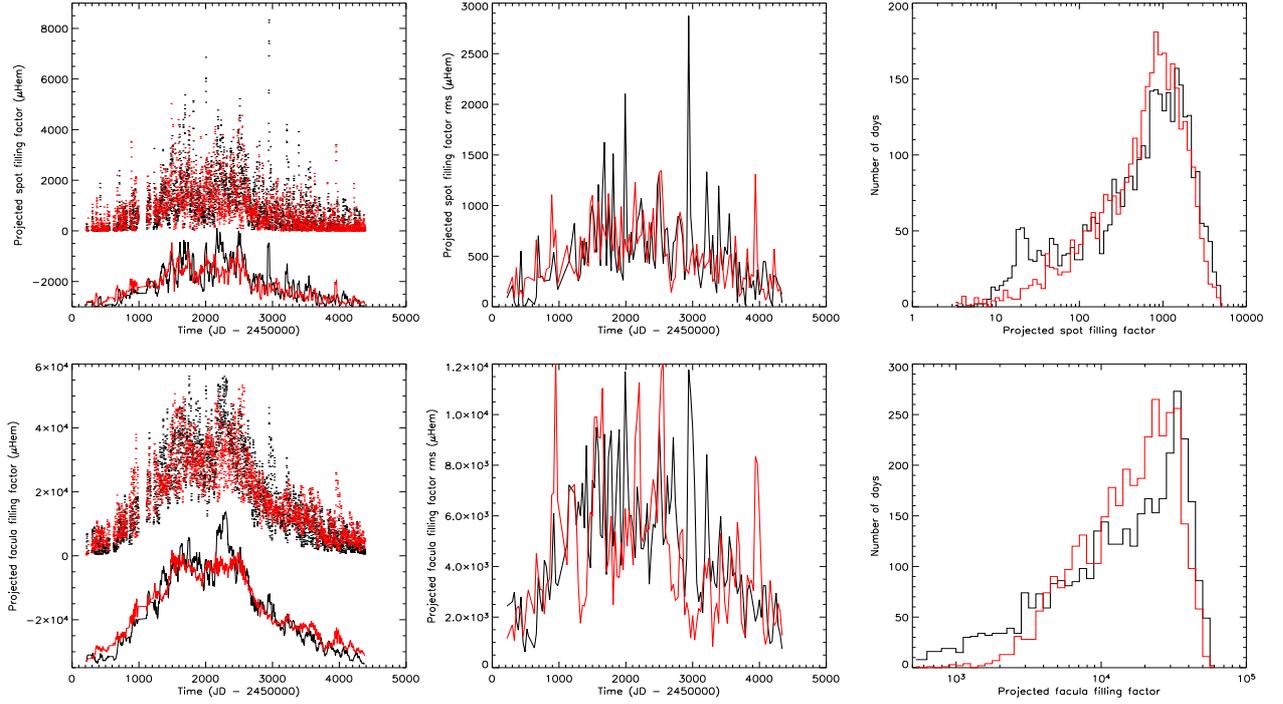}
\caption{\footnotesize{Projected spot ({\it top}) and bright feature ({\it bottom}) filling factors (in $\mu$Hem), for observed ({\it black}) and simulated ({\it red}) structures. {\it Left}: projected filling factor time series ({\it dots}) and averaged over 30 days ({\it solid lines}). Note that a vertical offset has been put on the averaged curves for visibility.  {\it Middle}: Rms of the projected filling factor over 30-day intervals. {\it Right}: histogram of the projected filling factor.}}
  \label{fillfactor}
   \end{figure*}

The solar observations are not made exactly in the plane of the solar equator, and the active structure distribution seems to be slightly shifted towards one visible solar hemisphere or the other, depending on the observation time, due to the noise level in the observed magnetograms\footnotemark. 
\footnotetext{As the noise level in the MDI magnetograms increases from the center to the edges of the images, the detectability level of a given structure is different between the two hemispheres, depending on its latitude.}
This effect is also visible on the bright feature latitude histogram, where the observed distribution is spread a little more towards higher latitudes.

\subsubsection{Filling factors}\label{ffdist}

Moreover, we compare the filling factors of the projected active structures over the full solar cycle for the observed and simulated patterns in Fig.~\ref{fillfactor}. We find the observed and simulated filling factor time series to be in very good agreement. The main difference between the observed and simulated distributions is located during the high-activity period: the activity peak present in the observations between approximately JD 2452200 and JD 2452400 is not well reproduced in the simulations. This difference does not come from a global lack of active structures in the high activity period of our model because {\it i}), the simulated daily spot number comes from Wolf number observations and {\it ii}), the facula distribution directly depends on the spot one (Sect.~\ref{listparam}). It rather originates in an occasional concentration of a few great structures in the observed pattern that is not reproduced in our model. Thus, this discrepancy is not a bias of our model but can be rather considered more as statistical noise. In the case of the bright features only, we also note a slight discrepancy between the observed and the simulated filling factors during the low activity period (the simulated filling factor being higher than the observed one). We attribute this to the larger number of very small bright features injected in our model (see before).

\begin{figure}[ht!]
   \centering
   \includegraphics[width=0.9\hsize,height=0.65\hsize]{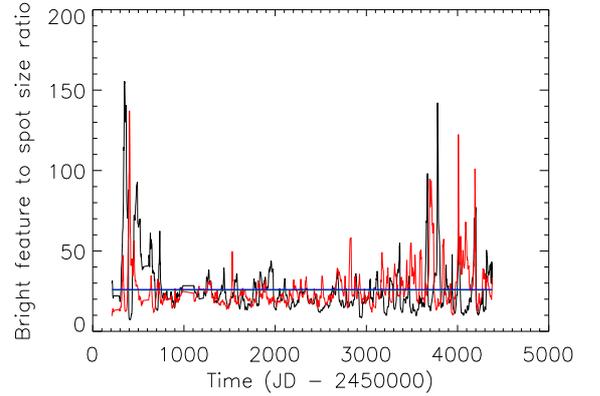}
  \caption{Facula-to-spot size ratio averaged over 30 days, for observed ({\it black}) and simulated ({\it red}) structures. {\it Blue solid line}: average value over the cycle.}
   \label{size_ratio}
   \end{figure}

Finally, we compare the ratio of the bright feature and spot filling factors over the solar cycle in Fig.~\ref{size_ratio} to estimate the evolution of the facula-to-spot size ratio. We find the ratios for the observed and simulated time series to be in very good agreement. One can also note that the size ratio reaches high values (from 50 to 150) during the low activity period (\ie~at the beginning and at the end of the solar cycle), whereas it is much lower (around 20) during the major part of the cycle. This is because during periods of low activity, the number of dark spots is very low (equal or close to zero), whereas there is always a ``background noise'' due to bright features, and especially the network. 

We find the average value of the size ratios over the total cycle to be very close to each other, with values of 25.7 and 26.0 for the observed and simulated structures, respectively. These values are in agreement with the studies of \cite{chapman97,chapman01,chapman11}. We also remark that the size ratio time series displayed by \cite{chapman11} look very much like ours over cycle 23, with peaks above 100 during low activity periods and minima around 25 during high activity ones. 

   \begin{figure*}[ht!]
   \centering
   \includegraphics[width=0.9\hsize,height=0.25\hsize]{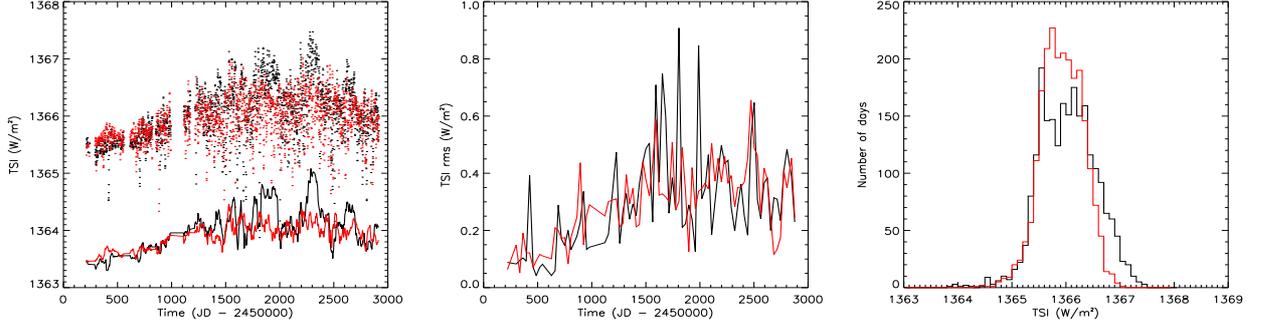}
   \caption{{\it Left}: observed TSI from \cite{frohlich98} ({\it black}) and reconstructed photometric contribution from all simulated structures ({\it red}). The photometric time series ({\it dots}) are displayed as well as their average over 30 days ({\it solid line}), with a vertical offset on the latter for visibility. {\it Middle}: Rms of the photometric time series over 30-day intervals. {\it Right}: Histograms of the photometric time series.}
   \label{phot}  
   \end{figure*}

Overall, we can conclude that our simulated solar activity pattern matches well the observed one and is thus a reliable model of a solar-like star magnetic activity. However, to definitely establish its validity, we compare the resulting RV and photometric time series. In particular, we check the impact of the differences between the two patterns on the time series.

\section{Comparison with the Sun observations}\label{results}
In this section, we first describe the way we simulate the time series for the different observables and how we take into account the stellar properties (limb-darkening, active structures temperature contrasts, impact on convective blueshift, physical and geometrical stellar properties). We then compare the RV time series simulated in the case of an edge-on solar-like star to the RV time series obtained with the observed activity pattern described in Paper II to establish the validity of our model.
\subsection{Description of the simulations}

\subsubsection{Limb-darkening}\label{limbd}
In our previous papers, the center-to-limb darkening was linear with respect to $\mu$ \citep[with $\mu = cos(\theta)$, $\theta$ being the angle to the center of the solar disc,][]{desort07}, as follows:
\begin{equation}
I(\mu) = 1 - \epsilon + \epsilon \ \mu
\end{equation}
with $\epsilon = 0.6$. In this case, the limb-darkening was not temperature-dependent and therefore applied to both the inactive photosphere and the active structures indiscriminately. Here, we decided to change our limb-darkening law, so as to: {\it i)} have a more accurate one; and {\it ii)} use a temperature-dependent one, that could be adaptable to different types of stars. We took the non-linear limb-darkening law from \cite{claret03}: 
\begin{equation}
\frac{I(\mu)}{I(1)} = 1 - \sum_{k=1}^{4} a_{k}(1-\mu^{\frac{k}{2}}) 
\end{equation}
where $I(1)$ stands for the intensity at the center of the stellar disc. The four Claret limb-darkening coefficients $a_{k}$ are the bolometric coefficients taken from ATLAS models \citep[see][]{claret03}. These coefficients are temperature-dependent. That is why such a law is better adapted to our simulation tool, as the effective temperature is one of the input parameters, and such a law can be extrapolated not only for the Sun, but for any star for which the effective temperature is known. It also means that the limb-darkening coefficients applied to the spotted stellar surface will be slightly different from the ones applied to the inactive photosphere. We therefore use a different limb-darkening for the photosphere and for the dark spots, depending on their respective temperature. As faculae show a very strong contrast variation depending on their position on the disc due to more complex processes, we directly define their contrast $C_{pl}(\mu)$ with respect to the limb-darkened photosphere (see below) without using the Claret law according to their temperature.
\subsubsection{Spot and bright feature contrast and photometric time series} 

To estimate the contrast of an active structure (dark spot or bright feature) with respect to the quiet photosphere, we use the procedure described in Paper II. This procedure is an independent, preliminary step to our simulations (\ie, to the generation of the spectra and of the RV and photometric time series), and is aimed at determining the active structure contrasts that we will use as input parameters. Starting from a given range of values for the spot and bright features contrasts, and based on observed structure patterns, we build (for each set of contrast values) two time series representing the respective relative contributions of dark and bright active structures to the stellar irradiance. A $\chi^{2}$ minimization between the sum of these contributions (quiet photosphere, dark and bright features) and the observed total solar irradiance (hereafter TSI) taken from \cite{frohlich98} is then performed over the solar cycle 23. We adopt the same value for the quiet Sun reference as in Paper II, \ie~$1365.46 \ W.m^{-2}$, which is very close to the average value found by \cite{crouch08} with their TSI model over twelve solar cycles.
Contrary to Paper II, we include in the procedure the influence of the center-to-limb darkening, so as to take into account the Claret limb-darkening law we will now use in our simulations. We end up with a spot temperature deficit $\Delta T_{sp} = -605 K$ and a facula contrast\footnotemark~$C_{pl} = 0.131618 - 0.218744\mu + 0.104757\mu^{2}$. We then use these contrasts as our input parameters in the simulations. As in Paper II, we finally compare the sum of the photometric contributions of spots and bright features obtained with our simulations with the observed TSI of \cite{frohlich98}, so as to check the validity of our activity model. The comparison is done on 2263 points, ranging from 1996 to 2003.

\footnotetext{As in Paper II, the spot contrast is a temperature contrast, while the bright structure (facula or network) contrast is defined as $C_{pl} \ = \ \frac{S_{pl}-S_{ph}}{S_{ph}}$, with $S_{pl}$ and $S_{ph}$ the facula/network and quiet photospheric irradiances, respectively.}

The irradiance obtained with our simulations and the observed TSI of \cite{frohlich98} are displayed in Fig.~\ref{phot}. We match quite well the observed irradiance, excepted for some peaks in the high activity period that are not reproduced in the model. This can be explained by the fact that our model does not reproduce well the occasional appearance of very large active structures or large active structure clusters (see Sects.~\ref{sizedist} and~\ref{ffdist}). Consequently, the distribution of the simulated TSI is slightly narrower than the observed TSI one, and the averaged rms of the simulated TSI is about $15\%$ lower in comparison (see Table~\ref{RVamp}). However, the temporal evolution of the simulated TSI rms matches well the observed one. As already stated, we consider that these differences do not affect the significancy of our simulations as a model.

\subsubsection{Attenuation of the convective blueshift}
For all following simulations, we adopt the same value for the attenuation of the convective blueshift as in Paper II, \ie~190~\ms. We refer to this paper for justification.

\subsubsection{Building the spectra}
We build the spectra from the input structure lists as described in \cite{desort07} and in Papers I and II. We assume a stellar mass of 1 \Msun, a temperature $T_{\rm eff}$ of 5800 K and a rotational velocity of 1.9 \kms~at the equator. We keep in this section the stellar inclination $i$ to 90\degr, corresponding to a star seen edge-on. Briefly, we use a synthetic spectrum from Kurucz models \citep{kurucz93} corresponding to a G2V star and apply it to each cell of the visible stellar 3D hemisphere divided into a grid. The spectrum is shifted to the cell radial velocity. In case of the presence of an active structure, it is weighted with a black-body law, taking into account the active structure contrast with respect to the quiet photosphere. To include the effect of the attenuation of the convective blueshift, the cell spectrum is redshifted by 190 \ms. This attenuation has other effects on the spectral lines, however \cite{dumusque14a} show that only considering the RV shift is sufficient to estimate the RV effect of active regions. 

All cell spectra are then balanced by the cell's projected surface and limb-darkening. We finally sum up all cell contributions to obtain the stellar spectrum. 

\subsubsection{Computation of the RV time series}
As in Paper II, we compute the RV using our Software for the Analysis of the Fourier Interspectrum Radial velocities \citep[SAFIR, see][]{galland05} on the built spectra as if they were actual observed spectra. We use only the wavelength range corresponding to the order $\sharp~31$ of the \harps~spectrograph, as done in Papers I and II. We obtain three distinct time series: two due to the respective photometric contributions of dark spots and bright features (hereafter the spot and facula time series, respectively); and the third due to the partial inhibition of the convective blueshift in the active structures (hereafter the convection time serie). For the latter, we consider only the inhibition of the convective blueshift (\ie~this effect is not weighted by the active region flux). We sum the three time series to obtain the total RV variations. Here we consider this sum to be a good approximation of the real RV variations given that the convective blueshift is dominant (see below).

   \begin{figure}[ht!]
   \centering
   \includegraphics[width=0.95\hsize]{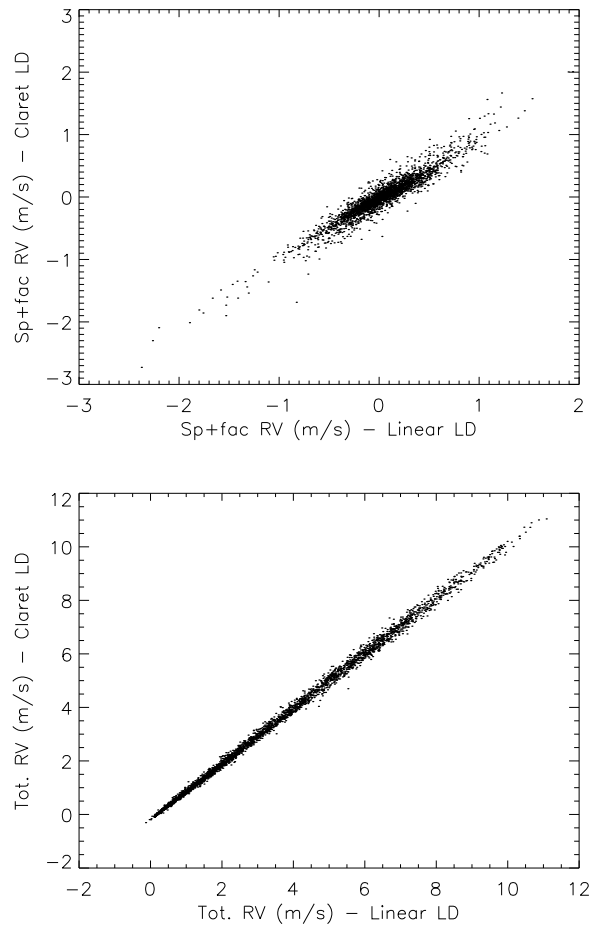}
   \caption{RV time series computed with Claret limb-darkening and corresponding structure contrasts versus RV time series computed with linear limb-darkening (as in Paper II). {\it Top}: contribution of spots and faculae. {\it Bottom}: sum of all contributions.}
   \label{obsobsnew}
   \end{figure}

\subsection{Validation of the activity model}

\subsubsection{RV time series based on the observed patterns}\label{sectcompobs}

In this section, we first check if the change of limb-darkening law and the corresponding change of the active structure contrasts have a significant impact on the resulting RV time series. To do so, we compute new RV time series corresponding to the observed solar activity patterns used in Paper II, but this time with the new limb-darkening law and structure contrasts. We then compare these new RV time series to the ones obtained in Paper II, using the same temporal sampling. We find the RV time series to be closely correlated and thus in good agreement. We display in Fig.~\ref{obsobsnew} the RV time series computed with the Claret limb-darkening versus the RV time series computed with the linear limb-darkening. We find both the slope of the fits and the correlation of the RV time series to be close to 1, as well in the case of the spot+facula RV time series as for the total RV ones. Despite the simpler limb-darkening law and a different active structure contrast, the results shown in Paper II are still valid for estimating the RV effect.

   \begin{figure*}[ht!]
   \centering
   \includegraphics[width=0.95\hsize,height=0.95\hsize]{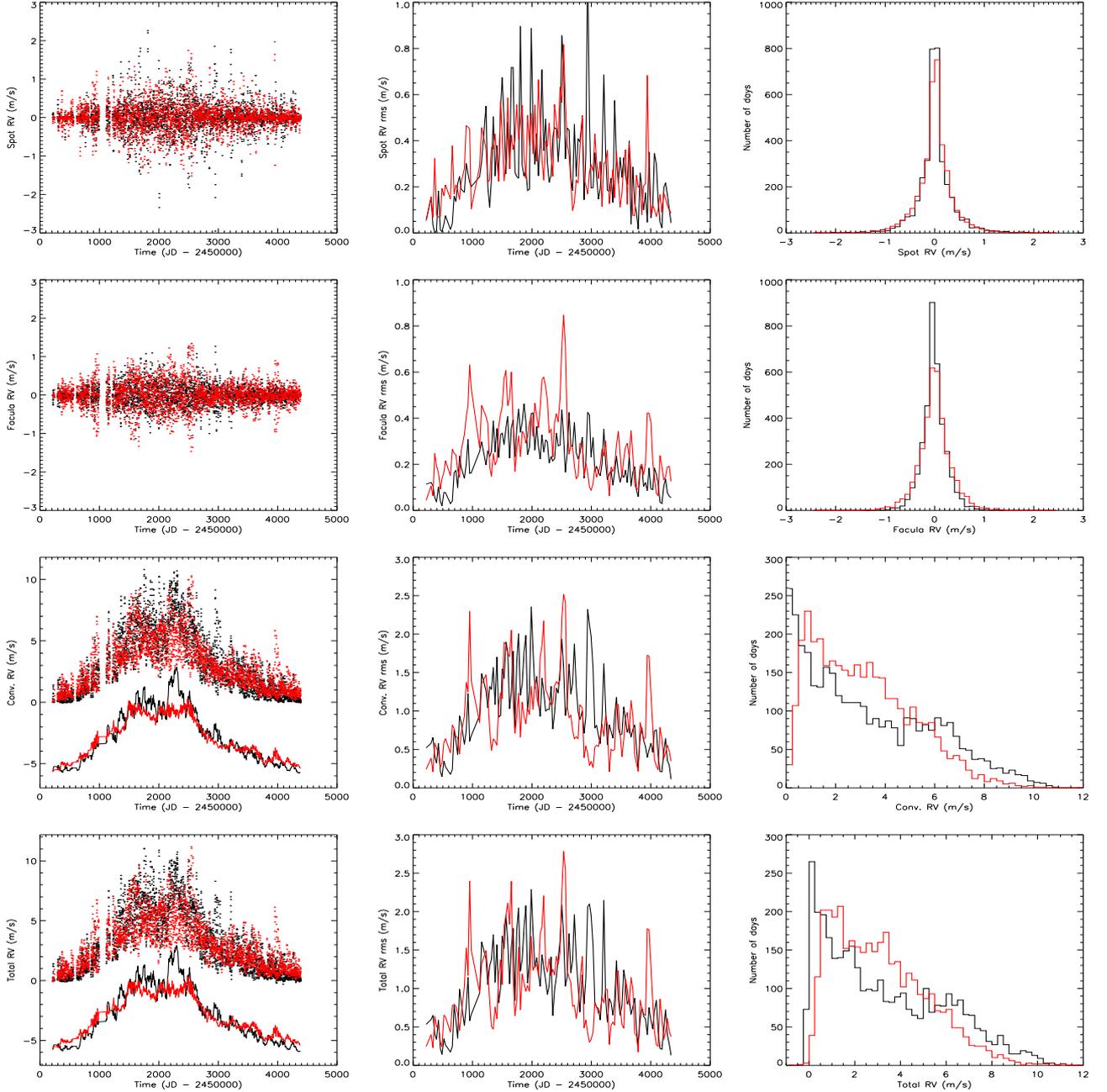}
   \caption{{\it From top to bottom}: Spot, facula, convection and total contribution to the RV time series, reconstructed from observations ({\it black}; same activity pattern as in Paper II) and simulated with our model ({\it red}). {\it Left}: RV time series ({\it dots}). The timescale is the one from Paper II. For the convection and total contributions, the RV time series averaged over 30 days are also displayed ({\it solid line}), with an offset for better visibility. {\it Middle}: Rms of the RV over 30-day intervals. {\it Right}: histograms of the RVs.}
   \label{rv}
   \end{figure*}

\subsection{Comparison between the RV time series based on the observed and simulated patterns}\label{comprvobssim}

We compare here the RV time series based on the observed and simulated activity patterns (both being computed with the Claret limb-darkening) so as to assess the validity of our activity model. We compare the RV time series for the different contributions (spot, facula, convection and all). All time series, as well as their dispersion over the cycle and their histograms, are displayed in Fig.~\ref{rv}. We find the amplitudes of the observed and simulated RV time series to be in very good agreement for the spot and facula time series, with very similar histograms. As for the RV time series corresponding to the convective component, it is closely related to the facula filling factor (with a Pearson correlation coefficient of 0.97) and widely dominates the total RV signal, as in Paper II. The main visible difference comes from the activity peak at around JD 2452200-2452400 in the RV reconstructed from observations, which is not echoed in the simulated RV. We already discussed the origin of this difference in Sect.~\ref{comps}. In the low activity period, the averaged amplitude of the simulated convection time series is about 20\% higher than for the observed one (see Table~\ref{RVamp}). We attribute this small discrepancy to the excess of very small bright features in the model that we discuss Sect~\ref{comps}.

  \begin{figure}[ht!]
   \centering
   \includegraphics[height=1.3\hsize]{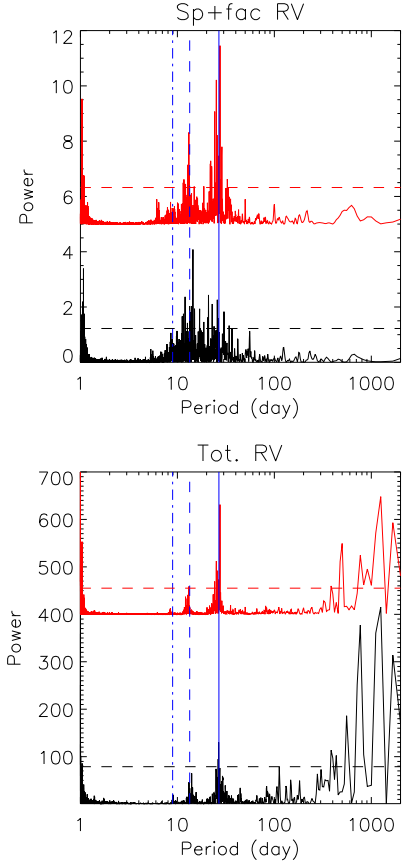}
   \caption{Lomb-Scargle periodograms of the RV time series based respectively on the observed ({\it black solid line}) and simulated ({\it red solid line}) activity patterns, alongside with the 1$\%$ false-alarm probabilities (FAP, {\it dashed lines}). {\it Top}: spot+facula contribution. {\it Bottom}: sum of all contributions. Note that in both panels the periodograms corresponding to the simulations are vertically shifted for visibility. The solar rotation period and its two first harmonics are displayed in blue ({\it solid}, {\it dashed} and {\it dashed-dotted blue lines}).}
   \label{comp_perio}
   \end{figure}

 To have a better understanding of the RV signature of our simulated activity pattern, we also display the RV rms computed over 30-day intervals over the cycle in Fig.~\ref{rv}. This gives an idea of the temporal evolution of the RV time series dispersion. We find the evolution of the RV dispersion during the solar cycle to be in good agreement for the observed and simulated time series. 
We also provide the RV rms for the different components taken over the complete cycle and for low and high activity periods in Table~\ref{RVrms}, and the RV amplitudes in Table~\ref{RVamp}. The low and high activity periods are the same as in Paper II. We note that:
\begin{itemize}
\item For the spot and facula simulated time series, the rms is slightly higher than for the observed time series in the low activity period (about two times higher in the case of bright features). However it is not the case for the convective time series. 
\item In the case of the convective component, the rms over the complete cycle is about 24\% lower for the simulated data than for the observed data. When looking at the temporal evolution of the RV rms displayed Fig.~\ref{rv}, we can see that this discrepancy mostly comes from a few high activity peaks in the observed time series.
\end{itemize}
We finally compare the Lomb-Scargle periodograms of the RV time series based on the observed and simulated activity patterns. The periodograms are displayed in Fig.~\ref{comp_perio} in the case of the ``photometric'' contributions (spots and faculae) and of the sum of all contributions. The respective periodograms for the observed and simulated patterns remarkably present the same characteristics: 
\begin{itemize}
\item For the ``photometric'' component, the power is mostly concentrated at the solar rotation period and its two first harmonics.
\item For the sum of all contributions, there is still power at the rotation period, but there is much more power at longer periods (between 300 and 1000 days) due to the long-term activity cycle.
\end{itemize}

\begin{table}[t!]
\caption{\footnotesize{RV rms (in m/s) for the different components.}}\label{RVrms}
  \begin{center}
  \begin{tabular}{c c c c c c}
  \hline
  \hline
Simulated pattern  & spots & faculae & sp+fac & conv. & total       \\
  \hline
All &  0.34    &  0.32  & 0.33   &  1.98     &  2.00     \\
Low &  0.17    &  0.16  & 0.14   &  0.48     &  0.51     \\
High & 0.47    &  0.38  & 0.47   &  1.36     &  1.52     \\
  \hline
Observed pattern\tablefootmark{\dagger} & spots & faculae & sp+fac & conv. & total  \\
  \hline
All & 0.37    &   0.25  & 0.32   &  2.59     &  2.62     \\
Low & 0.10    &   0.08  & 0.07   &  0.47     &  0.47     \\
High& 0.52    &   0.35  & 0.42   &  1.50     &  1.53     \\
  \hline
\end{tabular}
\tablefoot{
\tablefoottext{\dagger}{\ie~structure pattern from Paper II and Claret non-linear limb-darkening.}
}
\end{center}
\end{table}
 
\begin{table}[t!]
\caption{\footnotesize{RV rms and peak-to-peak amplitude (in m/s) and relative rms photometry for the three periods.}}\label{RVamp}
  \begin{center}
  \begin{tabular}{c c c c }
  \hline
  \hline
Simulated pattern  & rms RV & ampl RV & rms phot   \\
  \hline
All &  2.00    &  11.3     & $2.83 \ 10^{-4}$   \\
Low &  0.51    &  2.5      & $1.06 \ 10^{-4}$       \\
High & 1.52    &  7.7      & $2.97 \ 10^{-4}$       \\
  \hline
Observed pattern\tablefootmark{\dagger} & rms RV & ampl RV & rms phot \\
  \hline
All              & 2.62    & 11.4   & $3.6 \ 10^{-4}$ \\
Low              & 0.47    & 2.1    & $1.2 \ 10^{-4}$ \\
High             & 1.53    & 8.4    & $4.5 \ 10^{-4}$ \\
  \hline
  \end{tabular}
\tablefoot{
\tablefoottext{\dagger}{\ie~structure pattern from Paper II and Claret non-linear limb-darkening.}
}
\end{center} 
\end{table}

 Nonetheless, we note a significant difference: for the spot+facula RV signal, the peaks at half the rotation period are emphasized compared to the peaks at the rotation period in the case of the observed pattern, whereas it is the contrary for our simulated pattern. We already noticed and discussed in Paper II the predominance of the power at half the rotation period in the RV time series derived from observed solar patterns. We explained it by the presence of two symmetrically active longitudes on the solar surface. Since here we do not impose two active longitudes separated by 180$\degr$ over the whole time series but rather variable active longitudes, this effect may not be very important. An alternative explanation could be that the difference between the periodograms corresponding to observed and simulated patterns may originate in the active structure quick growing phase (which is at present not reproduced in our model). 

   \begin{figure}[ht!]
   \centering
   \includegraphics[width=0.7\hsize,height=1.2\hsize]{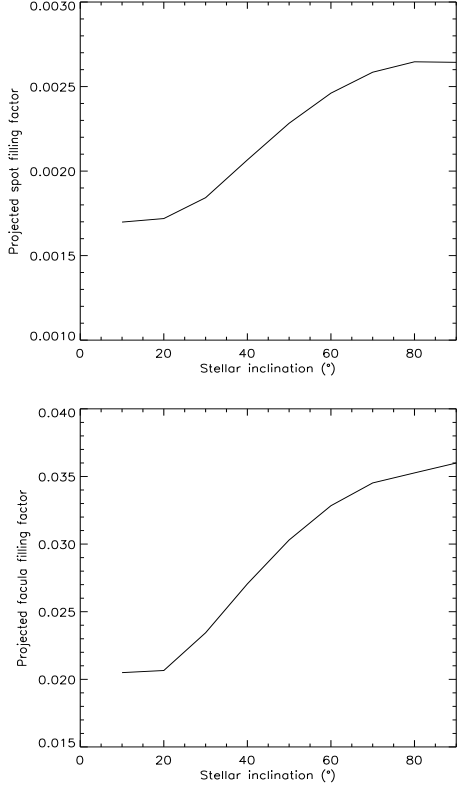}
   \caption{{\it Left}: projected spot filling factor at maximum activity (in fraction of the stellar surface) versus stellar inclination {\it i}. {\it Right}: same for bright features.}
   \label{varfp}
   \end{figure}

This may also explain the smaller RV and photometric rms when compared to the observations.\\

As detailed above, small discrepancies in amplitude and dispersion on short timescales are present both in the photometry and in the RV between the time series derived from observed and simulated activity patterns. We consider that these differences likely originate in two sources: first, the spot size distribution, where the few larger spots are hard to model; and second, the active structure growing phase. Overall, we consider that the temporal evolution over the complete cycle of both the amplitude and dispersion of the simulated time series match well the time series derived from the observed activity pattern. Hence, we conclude that: {\it i}) these short-term discrepancies are not significant enough to question our model reliability; and {\it ii}) the comparison between the RV time series derived from observed and simulated patterns then assess the overall validity of our model.

\section{Photometric and RV time series of inclined solar-type stars}\label{inclin}

   \begin{figure}[ht!]
   \centering
   \includegraphics[width=0.7\hsize,height=1.2\hsize]{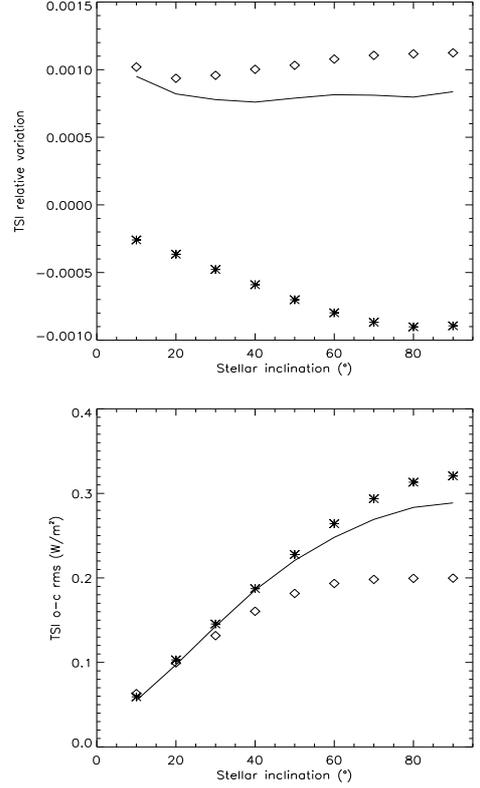}
   \caption{\footnotesize{{\it Top}: Relative variation of the TSI over the activity cycle (\ie, peak-to-peak amplitude of the TSI time series averaged over 30-day intervals, expressed in fraction of the quiet Sun irradiance reference) versus inclination {\it i} ({\it solid line}). The respective contributions of spots ({\it stars}) and bright features ({\it diamonds}) are also displayed. {\it Bottom}: short-term TSI dispersion (rms of the residuals after substraction of a 30-day averaged time series) versus {\it i}.}}
   \label{varphoto}
   \end{figure} 

For solar-type stars of the same age as the Sun, it is commonly assumed that active structures are mostly concentrated in a belt around the stellar equator (even if bright features are much more dispersed in latitude). A different inclination of the stellar rotation axis should then have a significant effect on the various time series that needs to be investigated. In this section, we perform the same simulations as above, but for different inclinations of the stellar rotation axis. We consider inclinations between $i = 10\degr$ (\ie, for a star seen nearly pole-on) and $i = 90\degr$ (star seen edge-on), with a sampling of 10\degr. The time series are studied with their original temporal sampling, \ie~4566 days and no gaps. 

\subsection{Photometric time series}\label{phot_serie}


The impact of the inclination {\it i} of the stellar rotation axis on the long-term solar irradiance variations was studied by \cite{knaack01}. The authors computed the solar irradiance corresponding to a 3-component model (quiet Sun, dark spot and bright facula) at activity extrema and and for a variable inclination. As for their active region distributions, they used simple active latitude belts with contrast corresponding to the given active structure (\ie~no individual structure was introduced in their model). According to them, the apparent active structure surface coverage decreases with a decreasing {\it i}. They nonetheless expected an increase of the TSI with a decreasing {\it i} since bright features are limb-brightened while the contrast of dark spots is roughly independent of the limb-darkening. Thus, when going from an edge-on toward a pole-on configuration, the impact of dark spots would decrease following their apparent covered surface. On the contrary, the decrease of the apparent surface covered by bright features would be at least compensated by their increased contrast since we would see them mainly on the limb. We display in Fig.~\ref{varfp} the evolution of the spot and facula projected filling factors with {\it i}. We find the projected spot filling factor to decrease by about $35\%$ when going from edge-on ({\it i}$= 90\degr$) to nearly pole-on ({\it i}$= 10\degr$), and the projected facula filling factor to decrease by about $43\%$. These results are quite similar to those found by \cite{knaack01}. We note that the filling factor does not tend towards 0 for $i \simeq 0\degr$.\\

 The relative variation of the TSI during the cycle (which is of the order of $0.1\%$ of the quiet Sun irradiance) vs. {\it i} is displayed in Fig.~\ref{varphoto} (upper panel). We find it to increase by only $14\%$ when going from {\it i}$= 90\degr$ to {\it i}$= 10\degr$. This is much smaller than the $40 \pm 10 \ \%$ increase predicted by \cite{knaack01} with comparable input parameters but a much simpler model. Yet, it points towards the same verdict as pulled through by \cite{knaack01}, \ie~that photometric variations of seemingly inactive Sun-like stars cannot be explained by an inclination effect. We conclude that for a solar-like star with a similar activity configuration, the effect of an inclined rotation axis on the long-term variations of the total irradiance (of the order of the solar cycle length) is relatively small. On the contrary, the TSI short-term variations \citep[which were indeed not studied by][]{knaack01} are strongly impacted by the inclination. As illustrated in Fig.~\ref{varphoto} (lower panel), we find the TSI short-term dispersion to be decreased by a factor $\sim 6$ when going from {\it i}$= 90\degr$ to {\it i}$= 10\degr$. A possible explanation for such a decrease is the following:
\begin{itemize}
\item First, for smaller inclinations, we mainly see the effect of bright features as they are more extended towards the higher latitudes than the dark spots.
\item Then, due to the activity configuration (where active structures are mainly located on two belts on both sides of the solar equator), for small inclinations we see the same structures during all the rotation period and not during half a period. 
\end{itemize}
Therefore, in the case of a star seen nearly pole-on, the short-term dispersion of the irradiance (of the order of a few rotation periods) originates no more in the structure crossing of the visible hemisphere for each half rotation period. It originates only in the structures appearance and decay. To confirm it, we display in Fig.~\ref{planche} the Lomb-Scargle periodograms of the simulated TSI for representative inclinations. For a star seen edge-on, the periodogram is dominated by power at the rotation period of the star. When going towards smaller inclinations, the signal at the rotation period gradually decreases until it disappears completely for a star seen nearly pole-on. On the contrary, the signal at a much longer period (which is likely induced by cycle-related long-term periodicities of the order of the cycle length) becomes increasingly preponderant for smaller inclinations. A reason for which the signals at long term periods (in the 600 to 1500-day range) become increasingly dominant with decreasing {\it i} in the TSI (and to a lesser degree in the total RV) periodograms could be the following: for nearly pole-on configurations, we see the active structures on one stellar hemisphere only, whereas we see them on two hemispheres for configurations closer to edge-on. This may induce a kind of an averaging effect on the long-term signal for the edge-on configuration and explain its increase with decreasing {\it i}. We finally display in Fig.~\ref{photo1090} the simulated TSI for representative configurations. As we found above, the long-term amplitudes of the two time series are nearly the same; on the contrary the dispersion is widely reduced with {\it i}.

\subsection{RV time series}\label{RVINC}

    \begin{sidewaysfigure*}
    \centering
     \includegraphics[angle=90]{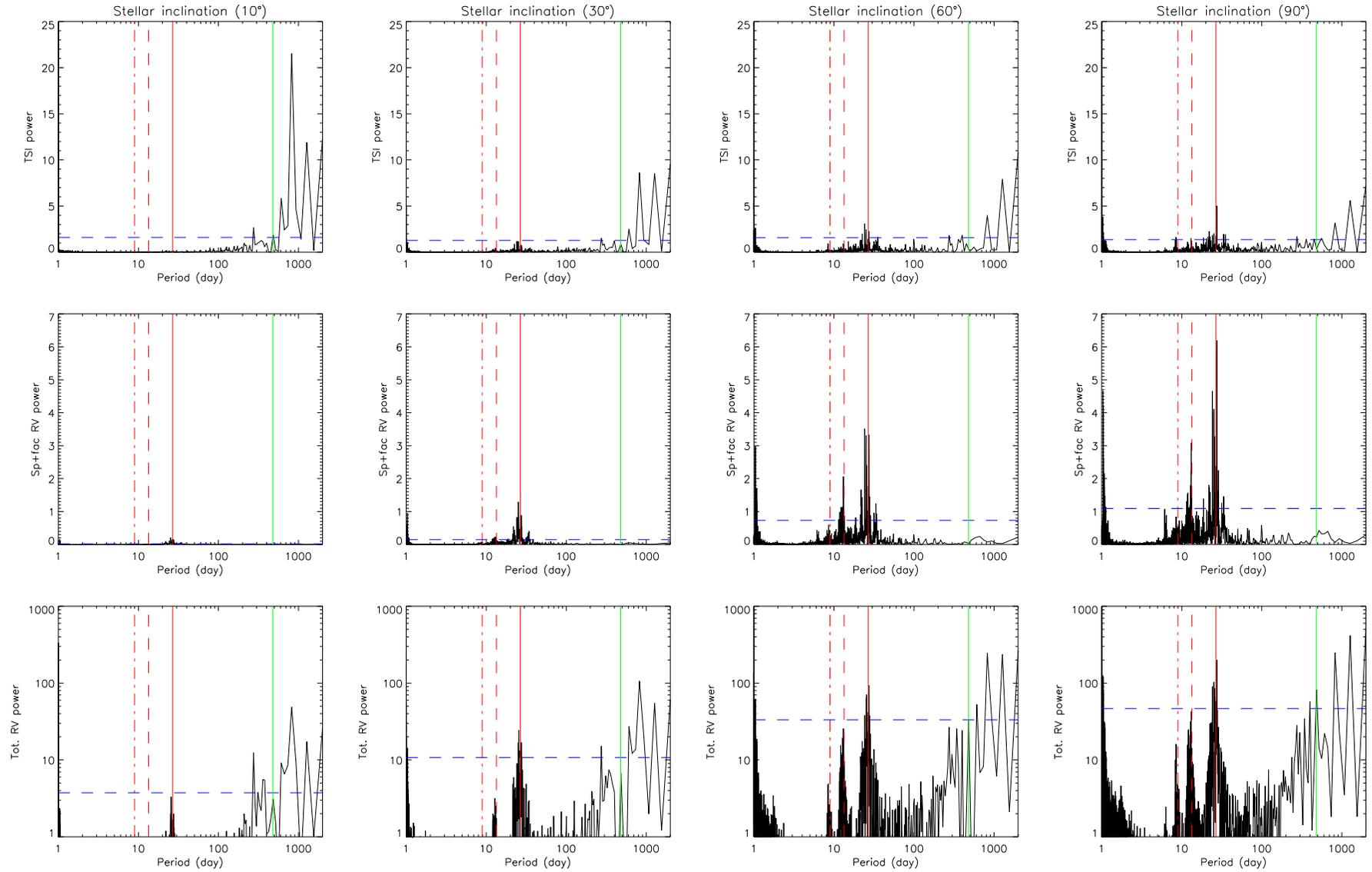}
     \caption{\footnotesize{{\it Top}: Periodogram of the TSI for significant inclinations ({\it from left to right}: i $=10\degr$, $30\degr$, $60\degr$ and $90\degr$). The $1\%$ false-alarm probabilities (FAP) are displayed ({\it blue dashed line}), as well as the equatorial stellar rotation period and its two first harmonics ({\it red solid}, {\it dashed}, and {\it dotted-dashed lines}, respectively), and the 480-day period for the computing of the detection limits ({\it green solid line}). {\it Middle}: the same for the RV ``photometric'' component (spot+facula). {\it Bottom}: the same for the total RV.}}
     \label{planche}
   \end{sidewaysfigure*}

We now study the impact of stellar inclination {\it i} on our simulated RV time series. Our first main result is that in contrast with the photometry, both the amplitude and the dispersion of the total RV decrease with a decreasing {\it i}. This is well illustrated in Fig.~\ref{rv1090} where we display the RV time series for all activity components and for significant inclinations. The amplitude as well as the dispersion (RV rms) of the total RV taken over the complete cycle are decreased by a factor $\sim 6$ when going from an edge-on to a nearly pole-on configuration. 

We characterize both the long-term and short-term variations of the RV signals to investigate deeper the impact of inclination. Our results are illustrated in Fig.~\ref{rvrms}. The ``long-term'' variations simply correspond to the signal taken over the complete stellar cycle. To study the short-term variations of the signal, we perform a running average of the RV time series with a smoothing window of 30 days and substract it to the original data. Then the long-term variations should primarily be affected by the global activity cycle, whereas the short-term variations will come from rotation-related effects.

\begin{enumerate}
\item {\it Peak-to-peak amplitude}: For the total RV and when going from $i=90\degr$ to $i=10\degr$, the amplitude decreases by $\sim 80\%$ over the cycle and by $\sim 85\%$ on the short term. When taking only the ``photometric contribution'' (\ie, spots and faculae) into account, the peak-to-peak amplitude is decreased by nearly $90\%$ on the long-term, and by $87\%$ on the short-term..
\item {\it Dispersion}: the total RV rms decreases by $70\%$ with inclination on the long term, and by $85\%$ on the short term. In the case of spots and faculae only, the decrease is the same on both timescales and is of nearly $85\%$. For a star with a solar-like activity pattern seen almost pole-on, a jitter of $\sim 0.5$ \ms~can be expected if the observation timescale is of the order of the activity cycle length, and a jitter of $\sim 0.2$ \ms~can be expected over a month.
\item {\it Ratio between ``photometric'' and convective components}: we also study the evolution of the relative contribution of the ``photometric'' (\ie, due to spots and faculae) component to the total RV signal with the inclination. Over the cycle, the ``photometric'' fraction of the RV is reduced by a factor $\simeq 2$ when going from $i=90\degr$ to $i=10\degr$ (\ie, the convective component is increasingly preponderant with a decreasing inclination). On the contrary, on the short-term the ``photometric'' relative contribution remains nearly constant with {\it i}, at a level of $\sim 0.3$. This value is at least two times larger than the value of the ``photometric'' fraction on the long-term (\ie, $\sim 0.16$ in the edge-on configuration). This confirms that the spot+facula component has mostly a short-term effect on the RV signal, of the order of the rotation period.
\end{enumerate}

We finally study the Lomb-Scargle periodograms of the RV time series for representative inclinations, first in the case of the sum of the flux contributions of dark and bright structures, and then when adding the convective component. The periodograms are displayed in Fig.~\ref{planche}. Starting from i $=90\degr$, the RV periodogram for the spot+facula signal is dominated by the stellar rotation. The power is concentrated at the rotation period and its two first harmonics, with about two times and six times less power for the first and second harmonics, respectively. The power is induced by the active structure crossing of the visible hemisphere during about half the rotation period (as the active structures are mainly located near and around the stellar equator). When {\it i} decreases, we begin to see the active structures on a time longer than half the rotation period, until we see them permanently for the nearly pole-on configuration. There is thus nearly no remaining component in the spot+facula RV periodogram for $i=10\degr$. This is in good agreement with the $90\%$ decrease in jitter already described.\\

Starting from the edge-on configuration, the total RV periodogram is dominated: {\it i}) by a serie of peaks at long period (200 - 1000 days) due to the long-term activity cycle; {\it ii}) by peaks at the stellar rotation period and its three first harmonics.

  \begin{figure}[h!]
   \centering
   \includegraphics[width=0.9\hsize,height=1\hsize]{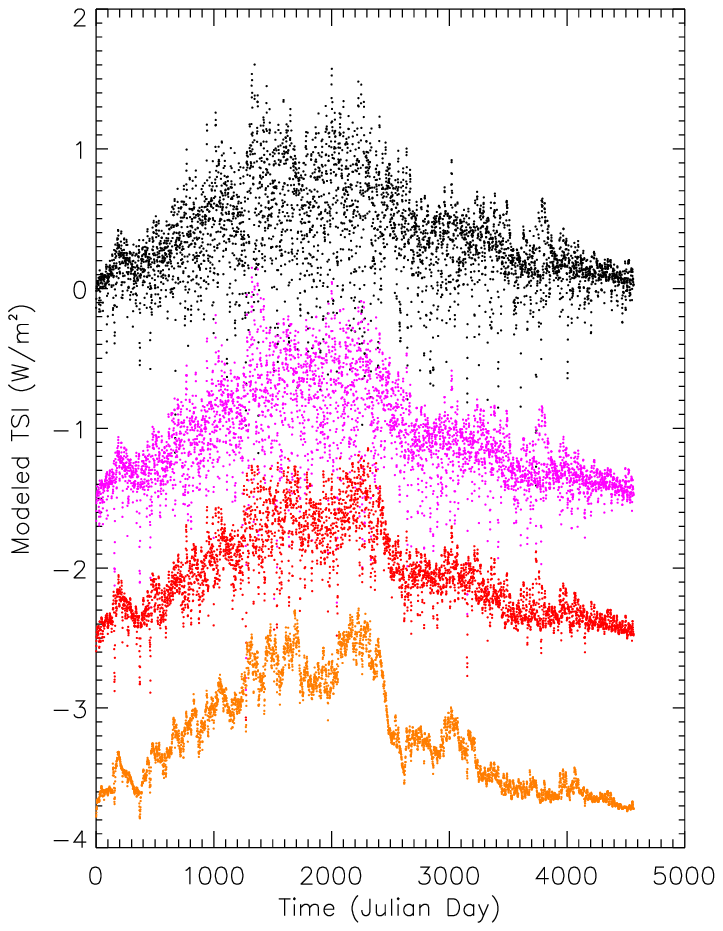}
   \caption{\footnotesize{Simulated TSI for: $i=90\degr$ ({\it black}), $i=60\degr$ ({\it purple}), $i=30\degr$ ({\it red}) and $i=10\degr$ ({\it orange}). Vertical offsets have been introduced for visibility.}}
   \label{photo1090}
   \end{figure}

  Expectedly, the peaks corresponding to the rotation period and its harmonics gradually decrease when going towards smaller inclinations, until they nearly disappear for $i=10\degr$. We also note that the harmonics of the rotation signal decrease significantly faster than the rotation signal itself. On the contrary, the long-term signal corresponding to the activity cycle decreases slowly when $i$ decreases, but remains widely above the $1\%$ false-alarm probabilities (FAP) for $i=10\degr$. In the edge-on configuration, the power is equally distributed between the rotation and the long-term signals, but the latter become preponderant for smaller inclinations. This is in agreement with the decrease of the ratio of the RV ``photometric'' component over the total RV signal on the long-term we show above. We conclude that in the case of a star with a solar-type activity pattern, the stellar inclination strongly impacts both the observed stellar irradiance, RV variations and active regions filling factor (and hence the observed chromospheric activity). Indeed, the short-term variations of these various observables strongly decrease with a decreasing inclination. Then, on short to medium timescales (from a few days to a year, or well under the activity cycle length), it will probably be not possible to distinguish between a solar-like star seen nearly pole-on and an inactive star. Provided that the time baseline is sufficient, on timescales of the order of the cycle length, the activity cycle should however still be detectable for small inclinations, but with a reduced amplitude. It would then not be possible to distinguish it from a nearly inactive star with a low-amplitude activity cycle.

   \begin{figure}[h!]
   \centering
   \includegraphics[width=0.9\hsize,height=1\hsize]{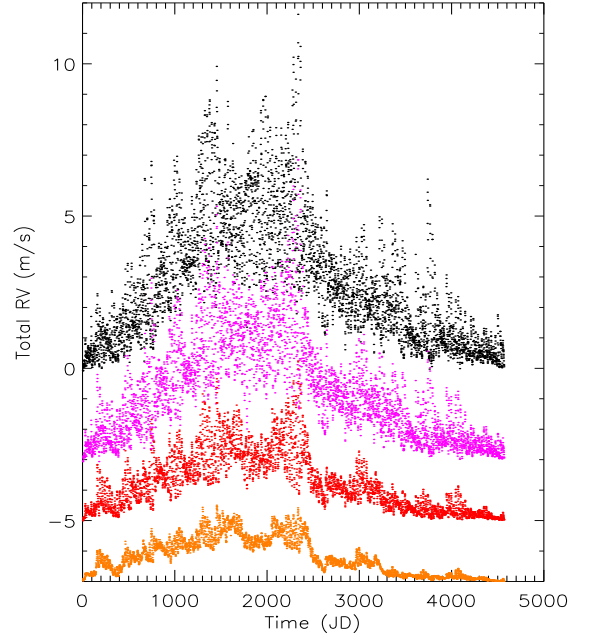}
   \caption{Simulated total RV for significant inclinations. {\it From top to bottom}: $i=90\degr$ ({\it black}), $i=60\degr$ ({\it purple}), $i=30\degr$ ({\it red}) and $i=10\degr$ ({\it orange}). Vertical offsets have been introduced for visibility.}
   \label{rv1090}
   \end{figure}

   \begin{figure*}[ht!]
   \centering
   \includegraphics[width=0.8\hsize,height=1.2\hsize]{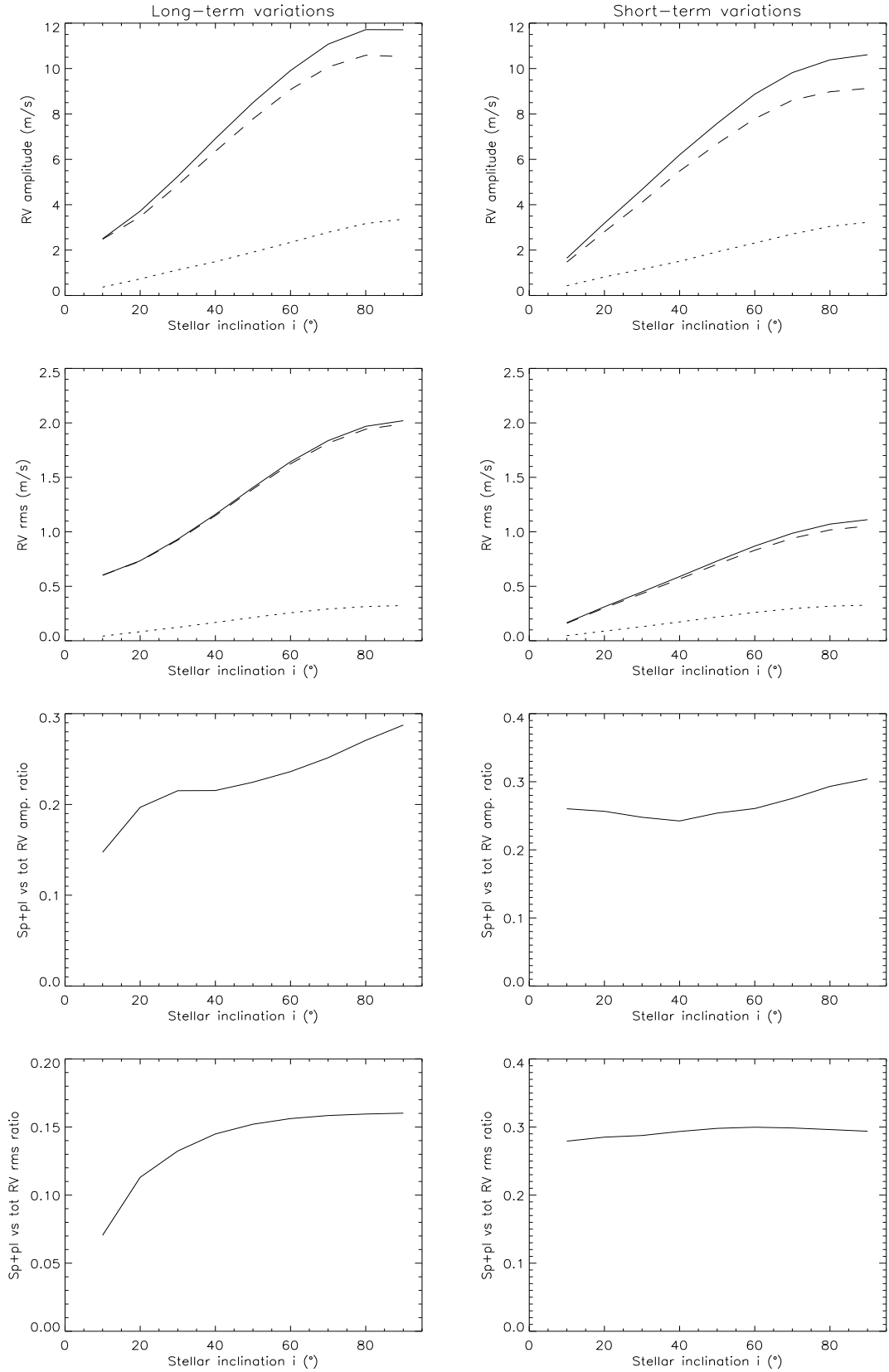}
   \caption{\footnotesize{Characterization of the RV signal versus sin(i). {\it Left}: Long-term variations of the signal (RV time series as produced by our simulations). {\it Right}: Short-term variations of the RV signal (residuals of the original RV data minus the 30-day averaged RV). {\it From top to bottom}: Peak-to-peak amplitude of the RV signal, RV dispersion (rms), ratio of the amplitudes of the spot+facula RV to the total RV, and ratio of the rms of these two signals as a function of sin(i). On the two top panels, the spot+facula ({\it dotted line}), convective ({\it dashed line}) and total RV ({\it solid line}) are also displayed as a function of sin(i).}}
   \label{rvrms}
   \end{figure*}

\subsection{Potential impact of stellar inclination on various correlations}

\subsubsection{Correlation between total RV signal and facula filling factor}
In Paper IV, we used the solar Calcium (Ca) index (S-index or log$(R^{'}_{\rm HK})$) as a tool to correct partially the activity-induced RV from its convective component (using the observed solar activity pattern). Indeed, the Ca index is correlated with the chromospheric plage filling factor and hence with the photospheric facula filling factor \citep[][found a linear dependence between the Ca index and the facula filling factor in the case of the Sun]{shapiro14}.

As already explained, in the case of a solar-like star seen edge-on, the RV signal is closely correlated to the facula filling factor (as the convective component dominates the RV variations, see Sect.~\ref{comprvobssim} and Figs.~\ref{fillfactor},~\ref{rv}). When going toward smaller inclinations, we note that the correlation gets even stronger, with a Pearson correlation coefficient going from 0.97 for $i=90\degr$ to 0.99 for $i=10\degr$. This is in agreement with the fact that the relative contribution of the spot+facula component to the RV signal decreases with {\it i}. It would finally mean that the Ca index correction method can be used regardless of the stellar inclination.

   \begin{figure}[ht!]
   \centering
   \includegraphics[height=1.1\hsize]{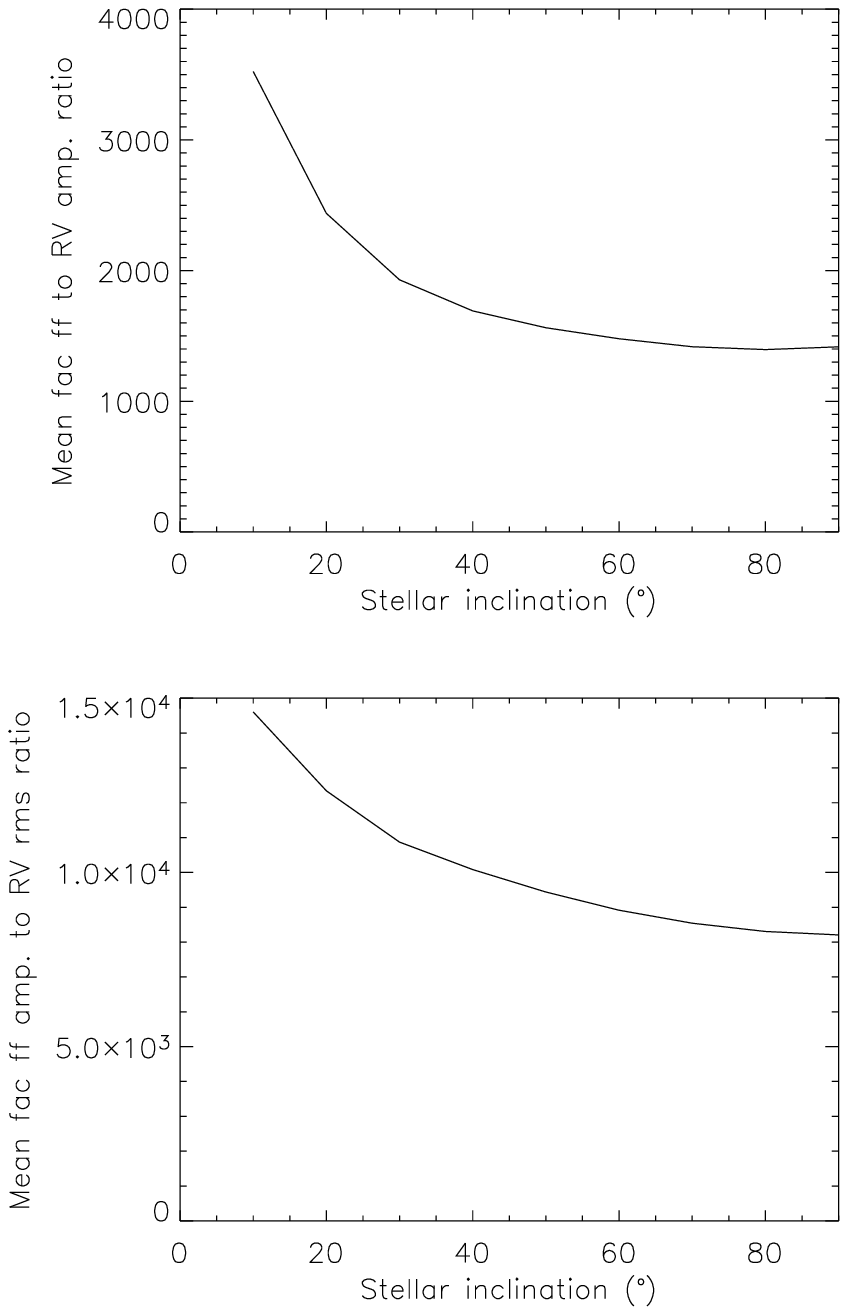}
   \caption{{\it Top panel}: Ratio of the mean facula filling factor over the total RV amplitude versus {\it i}. {\it Bottom panel}: Ratio of the mean facula filling factor over the total RV rms versus {\it i}.}
   \label{ratiofacRV}
   \end{figure}

\subsubsection{RV jitter, Ca index and facula filling factor}
In recent studies \citep{wright05,santos10,isaacson10,hillenbrand14}, the average level of RV jitter was commonly taken as a proxy for the stellar magnetic activity level, along with the Ca index. For different samples of FGK stars, the authors found loose correlations between the average RV jitter and the mean Ca index, generally with a significant amount of dispersion. In our simulations, we find that the ratio of the mean facula filling factor over the RV rms (calculated over the complete activity cycle) is not constant over the range of inclinations {\it i} we explored. We display in Fig.~\ref{ratiofacRV} the facula filling factor to RV amplitude and to RV rms ratios. We find both ratios to increase towards smaller inclinations (in agreement with Sects.~\ref{phot_serie} and~\ref{RVINC}, where the decrease of the RV amplitude and rms with {\it i} is more pronounced that the decrease of the facula filling factor with {\it i}). Provided that the Ca index evolves in the same way as the facula filling factor with {\it i}, this could partly explain the large dispersion in the (RV rms, mean Ca index) relation found in the studies cited above (as the stellar inclination remains unknown for most of the observed targets). It also means that a clear (RV rms, Ca) relation will remain hard to establish for future studies, unless the uncertainty on the stellar inclination can be removed.

\subsection{Detection limits}\label{thelimdets}

In this section, we compute the detection limits for the total RV time series over the range of stellar inclinations explored above.

\subsubsection{Approach}

 We use two different methods: the correlation-based method and the {\it local power amplitude} (hereafter LPA) method. Both were described in details and tested on real stellar RV data in \cite{meunier12}. In brief, the first method makes the correlation between the periodograms of a generated planetary RV signal (with the same temporal sampling as of the data) and of the actual RV data to which the planetary RV signal has been added (\ie, it determines the correlation of the power of the (stellar data + fake planet) periodogram vs. the power of the fake planet periodogram). The detection limit corresponds then to the minimal mass and period for which the correlation values are all above a given threshold, for 100 realizations spanning all orbital phases. The threshold (spanning here from 0.003 for $i$ = 10$\degr$ to 0.05 for $i$ = 90$\degr$) corresponds to the maximum of the correlations obtained for a very low mass planet (here $\sim$0.6 \ME. As for the LPA method, we compare the periodograms of the actual RV data and of a given generated planetary signal (for the same temporal sampling), but within a localized period range around the given planetary period. The detection limit at this period corresponds then to the minimal mass for which the maximum power of the planetary RV periodogram (within the limited period range) is always above the maximum power of the actual RV data periodogram in the period range.

 We compute here our detection limits for only one period of 480.1 days, corresponding to a separation of about 1.2 au, as in Paper II. It is a representative value of the outer boundary of the HZ for a solar-type star. The total time span is always fixed to the complete simulation time span (\ie~4566 days). As in Papers I and II, we assume the star to be observed eight months a year (meaning that we take into account only 2978 RV points instead of 4566). We assume different temporal samplings: all points (\ie~one-day sampling or 1:1), 1 point every 4 nights (1:4), 1 point every 8 nights (1:8) or 1 point every 20 nights (1:20); that is to say that the (1:20) series has 20 times less points that the (1:1). We also assume different precisions on the RV (no added noise, 0.01, 0.05 and 0.1\ms~noise levels), as we did in Paper I.

Finally, we compute the detection limits vs. inclination for two different cases:
\begin{enumerate}
\item  In the first case, we compute the detection limits considering that the hypothetical planet is always seen orbiting edge-on for all stellar inclinations (\ie, there is an increasing spin-orbit misalignment with decreasing inclination).
\item In the second case, we consider that the hypothetic planet always orbits in the stellar equatorial plane (\ie~there is always spin-orbit alignment). 
\end{enumerate}

Considering these two distinct cases is important as in most cases when searching for planets with RV, we do not know neither the stellar inclination nor the inclination of the planet orbit with respect to the line of sight. As for the inclination of the planet orbit, it is taken into account by giving minimal masses $m.sin(i)$ (with {\it i} denoting here the inclination of the planet orbit with respect to the line of sight) for detected planets. However, the detection limits are generally computed without knowing the stellar inclination itself, \ie~considering that the star is seen edge-on and that the activity-induced jitter is not reduced due to the projection effect. The detection limits we compute in case 1 correspond to this configuration and thus to the ``best-possible'' detection limits. As we go toward smaller stellar inclinations, they also correspond to more and more unlikely orbital configurations. Indeed, most of the exoplanets found so far indeed orbit in or near the stellar equatorial plane\footnotemark~\citep[even if in the specific case of Hot Jupiters, a significant fraction of systems show spin-orbit misalignments,~\eg][]{albrecht12,triaud14}. 

\footnotetext{http://exoplanets.eu}

  \begin{figure*}[ht!]
   \centering
   \includegraphics[width=0.83\hsize,height=1.22\hsize]{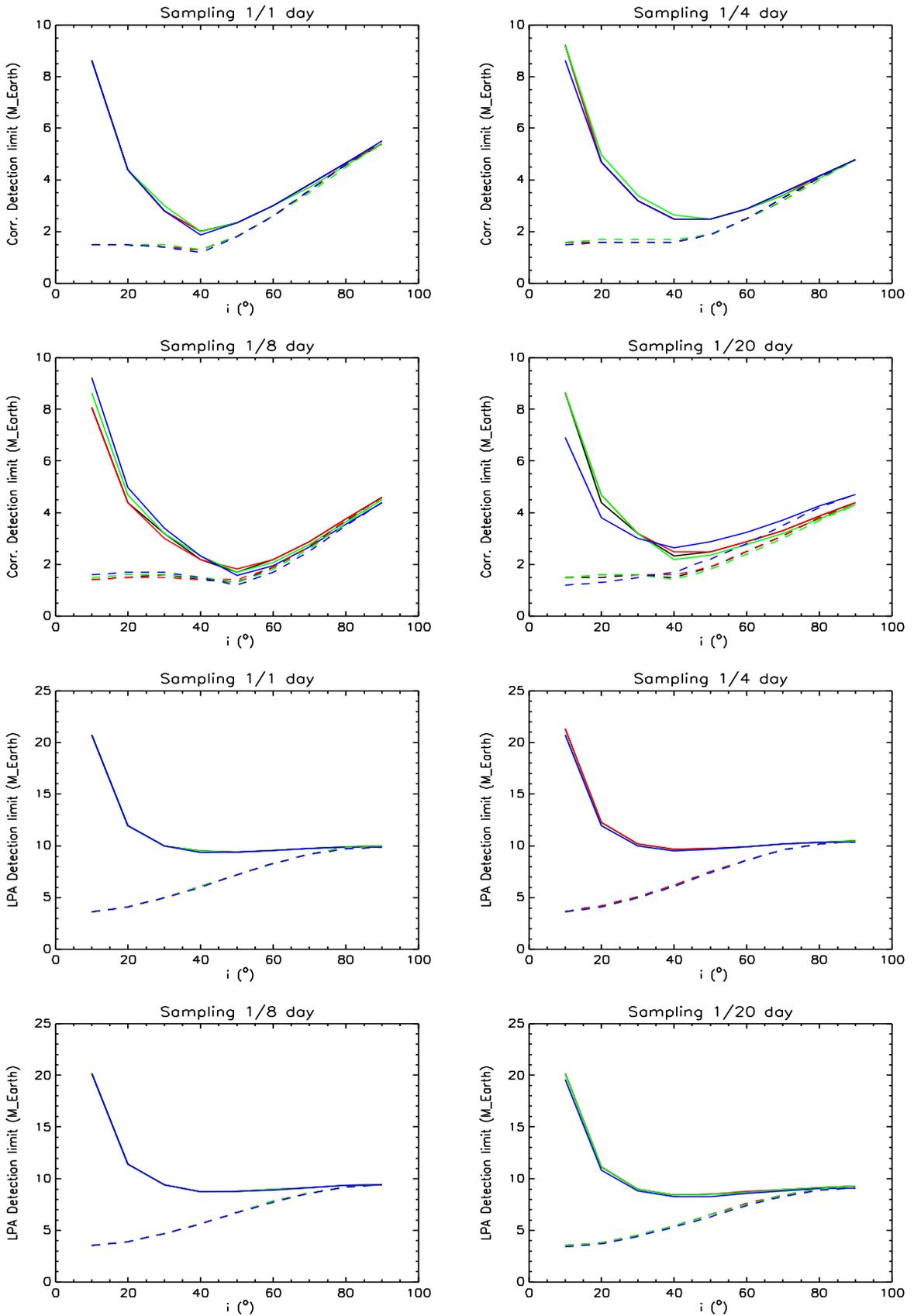}
   \caption{\footnotesize{Detection limits versus stellar inclination for the correlation-based ({\it four top panels}) and the LPA ({\it four bottom panels}) methods. The detection limits are displayed for different samplings and for different noise levels ({\it black}: no added noise, {\it red}: 0.01~\ms~noise, {\it green}: 0.05~\ms~noise and {\it blue}: 0.1~\ms~noise). {\it Solid line}: detection limits computed in case 2. {\it Dashed line}: detection limits computed in case 1.}}
   \label{limdets}
   \end{figure*}

On the contrary, the detection limits that we compute in case 2, \ie~when considering a spin-orbit alignment, correspond to a more conservative but more realistic assumption. The detection limits for the correlation-based and LPA methods are displayed in Fig.~\ref{limdets}.

\subsubsection{Results}

\paragraph{Comparison with previous results --} For the edge-on configuration, we can compare the detection limits (for the 1:1 sampling) to the detection limits computed in Paper IV\footnotemark, \ie~on RV time series derived from observed solar activity patterns, as the latter time series had a similar time span, sampling and number of data points. In the case of the correlation method, they are in a fairly good agreement (5.5~\ME~for the present time series versus 6.8~\ME~for the time series from observed patterns), whereas there is a certain discrepancy for the LPA method (9.9~\ME~and 15.7~\ME~for the simulated and observed RV time series, respectively).
\footnotetext{see Table 2 in Paper IV, case with no correction, total RV time series, Set 1.}

\paragraph{Comparison between the two methods --} The correlation-based method gives lower detection limits than the LPA method. This is also in agreement with our results from Paper IV. Note that we also compared the two methods in \cite{meunier12}, this time on actual stellar data. Out of a 10 target sample, we found in most cases that the LPA method gave lower detection limits. However in these cases, the targets were massive A-F Main-Sequence stars with medium to high \vsini~(from 7 to $\geq$ 175 \kms), some of them being young stars. They would then have a very different activity pattern than the solar one.

\paragraph{Impact of stellar inclination --} For the LPA method, the decrease of the detection limit with the stellar inclination {\it i} in case 1 (``best-possible'' detection limits) is consistent with the decrease of the activity-induced RV rms when going toward a smaller {\it i}. In case 2, we observe that the detection limits remain around 10 \ME~for $i\geq 40\degr$ before increasing toward higher masses for smaller inclinations. This should mean that for $i\geq 40\degr$, the decrease of the planetary-induced RV amplitude for a decreasing inclination of the system is counterbalanced by the decrease in the same time of the activity-induced RV jitter. For smaller inclinations, the decrease of the activity RV jitter is less pronounced and hence it becomes more difficult to detect the planet. This is consistent with the fact that the behavior of the activity-induced RV jitter with respect to {\it i} is not sinusoidal.\\

For the correlation method, the behavior of the detection limits is rather different. In case 1, the detection limits decrease with {\it i} until they reach a plateau at about 1.5 \ME~for $i\sim 40\degr$. When looking at case 2, surprisingly the detection limits begin to decrease with a decreasing {\it i}, to reach a minimum and best value of 2-2.5 \ME~for {\it i} $=40-50\degr$. For smaller inclinations, the detection limits increase with a decreasing {\it i}. Thus, the optimal configuration for the detection of the planet is an orbital plane inclined of $45\degr$ with respect to the line of sight (in the most probable case of spin-orbit alignment).

\subsubsection{Impact of the parameters}
In this section, we study in more details the impact of the main parameters for the computation of the detection limits (\ie, the added RV noise level and the temporal sampling. To do so, we first study their influence on the detection limits computed above (Fig.~\ref{limdets}). Then, we extend the study to a wider range of noise level or temporal sampling, but this time concentrating on one case ({\it i} $=$ 50$\degr$, case 2). This roughly corresponds to the inclination for which we get the better detection limits (in case 2).

  \begin{figure}[ht!]
   \centering
   \includegraphics[width=0.95\hsize,height=0.6\hsize]{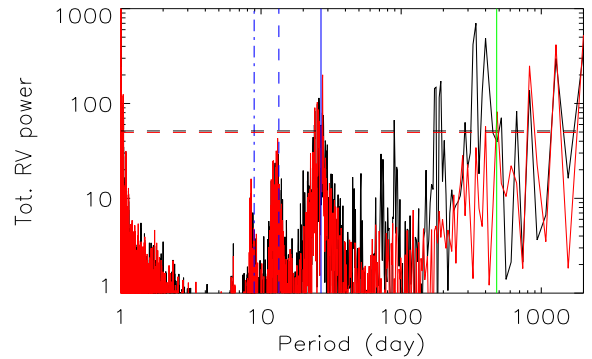}
   \caption{Tot. RV periodograms for {\it i} $=$ 90$\degr$, (1:1) sampling. {\it Black}: data taken eight months a year. {\it Red}: full data. The 1$\%$ FAP are displayed ({\it dashed lines}), as well as the solar rotational period and its two first harmonics ({\it blue}), and the detection limit period ({\it green}).}
   \label{compperiocut}
   \end{figure}

\paragraph{Temporal window --} Our detection limit computation is based on the periodograms of the RV time series taken eight months a year, so as to better mimic actual observations. The periodograms will thus be modified compared to the periodograms of the total RV time series taken over the full simulation time span, due to aliasing. We compare the periodograms of the total RV time series (in the {\it i} $=$ 90$\degr$ case) before and after removing 4 months a year in Fig.~\ref{compperiocut}. Removing four months a year indeed injects a large amount of power in the RV periodogram around one year ($\sim$ 360-400 days) as well as at lower harmonics (180 and 90 days). It will then impact our detection limit computation and deteriorate our detection limits, but it has to be taken into account to reproduce better the observations.

\paragraph{Temporal sampling --} We found in Paper IV that when taking into account all RV components, the detection limits got worse for the (1:20) sampling (\ie, for a largely degraded sampling), while they did not vary significantly when going from a (1:1) to (1:4) or (1:8) samplings. Fig.~\ref{limdets} shows that the LPA detection limits do not vary significantly when going from the (1:1) to the (1:20) sampling. As for the correlation detection limits, they do not vary significantly when going from the (1:1) to the (1:8) sampling, but get worse for the (1:20) case.\\

\begin{figure}[ht!]
   \centering
   \includegraphics{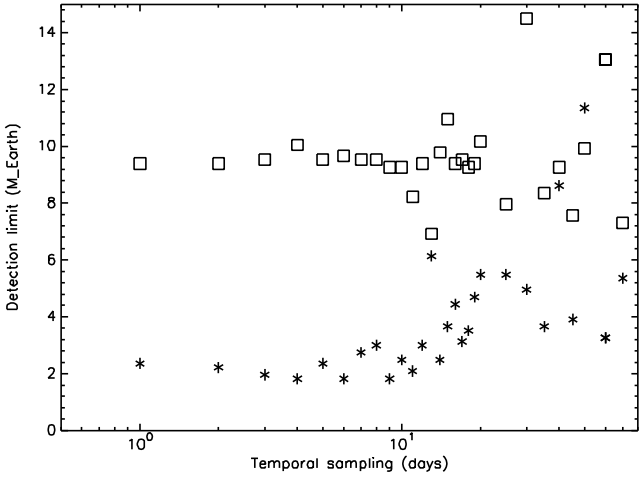}
   \caption{Detection limits vs. temporal sampling for {\it i} $=$ 50$\degr$, no noise. {\it Squares}: LPA method; {\it Stars}: correlation method.}
   \label{limdetsamp}
   \end{figure}

We now focus on the {\it i} $=$ 50$\degr$ case to study the sampling impact in more details. We compute the detection limits for an extended sampling range (from (1:1) to (1:70)) and display the results in Fig.~\ref{limdetsamp}. We find that both the correlation and the LPA detection limits are nearly independent from the temporal sampling up to the (1:10) sampling, and get worse for more degraded samplings (even if they show a large dispersion in the LPA case). The large dispersion at large temporal samplings probably reflects the larger uncertainty on the detection limits for a smaller number of points.

\paragraph{Added noise --} The detection limits computed with both the correlation and LPA methods are nearly independent from the added RV noise for a noise level up to 10 \cms~(Fig.~\ref{limdets}), \ie~for the best RV accuracy expected on future spectrographs. This is in agreement with the results of Paper IV. We consider that the small variations seen between the detection limits computed for the different noise levels in the case of the correlation method are not significant.\\

  \begin{figure}[ht!]
   \centering
   \includegraphics{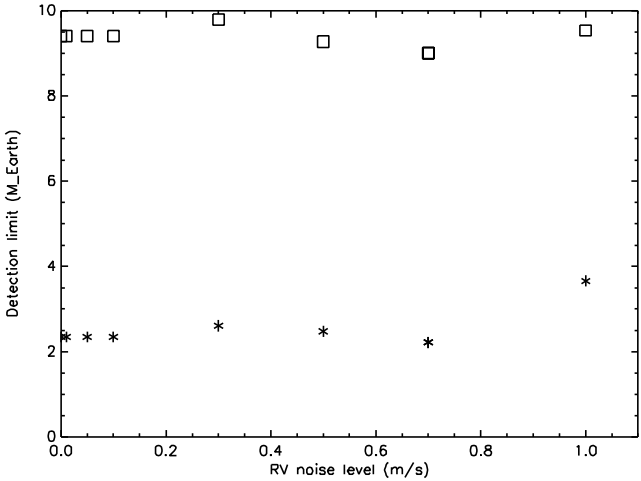}
   \caption{Detection limits vs. RV added noise for {\it i} $=$ 50$\degr$, (1:1) sampling. {\it Squares}: LPA method; {\it Stars}: correlation method.}
   \label{limdetnoise}
   \end{figure}

We study then the impact of the added noise for an extended range of noise levels, up to 1 \ms~(\ie, the current RV accuracy reached on the better spectrographs such as \harps), in the {\it i} $=$ 50$\degr$ case (Fig.~\ref{limdetnoise}). The LPA detection limits are constant up to 1 \ms, while the correlation detection limits remain constant up to 0.7 \ms~with a slightly larger value at 1 \ms.

\section{Conclusion and perspectives}\label{conclu}

We built a fully parametrized model of the activity pattern of a solar-like star, including the dark spots and the bright features (large faculae and network). The model includes about 30 parameters that account for the different activity scales and the active structure behavior (most of them being well constrained by the litterature), and has a daily timescale over a complete activity cycle. Using the same approach as in the previous papers, we deduced the corresponding RV and photometric time series, taking the inhibition of the convective blueshift into account in the case of the RV. The simulated activity pattern, as well as the time series, were compared to the work done in Papers I, II and III with data from solar observations. We found our model to be in remarkably good agreement with the previous data, thus assessing its validity. 
We then study the case of a solar-like star seen under different configurations so as to estimate the impact of the stellar inclination on the activity-induced jitter. Our results are the following:
\begin{itemize}
\item For the stellar irradiance, the stellar inclination {\it i} has almost no effect on the amplitude of the long-term irradiance variations. The decrease of the active structure filling factor with a decreasing {\it i} mostly counterbalances the enhanced contrast of the structures, in agreement with some previous studies \citep{knaack01,shapiro14}. On the contrary, the short-term jitter is reduced by a factor 6 when going from an edge-on to a nearly pole-on configuration. 
\item For the RV variations, the peak-to-peak amplitude as well as the rms are reduced by a factor 8 to 10 on both the long-term and the short-term. For a nearly pole-on configuration, the remaining RV jitter is about 0.6~\ms~when considering the whole cycle, and is about 0.2~\ms~on shorter timescales.
\item When computed over the whole activity cycle, the activity-induced periodograms show power mainly in two period ranges: first at the stellar rotation period and its two or three first harmonics (induced by the crossing of active structures on the visible stellar hemisphere), and then at much longer periods, induced by more complex cycle effects. When going toward smaller inclinations, the power at the rotation period is reduced and nearly disappears for a pole-on configuration. The long-term cycle effects become preponderant.
\item In the case of a solar-like star, the convective component widely dominates the RV variations. The ``photometric'' contribution of spots and bright features account for about $35\%$ of the RV jitter and decreases to about $15\%$ for a nearly pole-on configuration.
\item For a solar-like star seen in a pole-on configuration, the photometric and RV (as well as presumably the chromospheric) activity-induced variations are most probably not distinguishable from those of a less active or almost inactive star.
\item Finally, when considering a realistic orbital configuration (\ie,spin-orbit alignment), the optimal configuration for planet detection is a system inclined by about $45\degr$. In this case, the lowest detection limits reach planetary masses of about 2\ME~at 480 days, without applying a correction to the RV signal.
\end{itemize}

According to previous studies, solar-like stars show a great diversity of activity levels and properties \citep{schroder13}. The Sun itself is considered to be an average star, not particularly active but not quiet either. Being fully parametrized and validated for the solar case, our activity model now allows us to explore a wide range of activity parameters and stellar properties. In this paper, we have focused on solar-like activity and convection level stars; we expect stars with a lower activity level and/or stars with a lower convection level (such as K-type stars) to be much less affected and then to exhibit lower detection limits in the corresponding HZ. We also emphasize that activity correction methods (such as the Calcium index or the $H_{\alpha}$ line) are commonly applied to RV time series and already allow to reach detection limits at the \ME~level in the HZ (Paper IV). Optimized observation and reduction strategies are also promising \citep[such as averaging, see \eg][]{dumusque11b}. Despite not being the focus of our paper, improving such methods and strategies, as well as looking for new ones, is an extremely important question. Future instruments with very high RV accuracy may therefore be critical in implementing very efficient correction tools to extract low mass planet signals. Our goal is to be able to test and even reproduce the photometric and RV variations corresponding to each activity and stellar configuration. This should allow a better understanding of the magnetic stellar activity and open the way toward a better correction of the activity-induced stellar jitter. 

In Paper III, we derived the astrometric time series corresponding to the observed solar activity pattern. Our activity model also allows us to produce the astrometric time series corresponding to the simulated activity pattern (apart of the photometric and RV ones). We will present our results in astrometry in a separate paper. This is quite justified as upcoming instruments such as the Nearby Earth Astrometric Telescope \citep[NEAT, see \eg~][]{malbet12} should allow for the first time to search for low-mass planets around nearby FGK stars with astrometry.

\begin{acknowledgements}
We would like to thank our referee (X. Dumusque) for his very useful comments on the manuscript.
\end{acknowledgements}
\bibliographystyle{aa}
\bibliography{Simon_Biblio}

\end{document}